\documentclass{emulateapj}

\usepackage[utf8]{inputenc}
\usepackage[english]{babel}
\usepackage{hyperref}
\usepackage{natbib}
\usepackage{color}
\newcommand{\vc}[1]{\textbf{\em #1}}

\newcommand{\pder}[2]{\frac{\partial #1}{\partial #2}}

\begin{document}

\title{Statistics of Reconnection-Driven Turbulence}
\author{G. Kowal\altaffilmark{1,2}, D.~A. Falceta-Gonçalves\altaffilmark{2}, A. Lazarian\altaffilmark{3}, \& E.T.~Vishniac\altaffilmark{4}}
\shorttitle{Statistics of Reconnection-Driven Turbulence}
\shortauthors{Kowal et al.}
\altaffiltext{1}{Núcleo de Astrofísica Teórica, Universidade Cruzeiro do Sul, R. Galvão Bueno, 868 -- Liberdade, CEP: 01506-000, São Paulo, SP, Brazil}
\altaffiltext{2}{Escola de Artes, Ciências e Humanidades, Universidade de São Paulo, Av. Arlindo Béttio, 1000 -- Ermelino Matarazzo, CEP: 03828-000, São Paulo - SP, Brazil}
\altaffiltext{3}{Department of Astronomy, University of Wisconsin, 475 North Charter Street, Madison, WI 53706, USA}
\altaffiltext{4}{Department of Physics \& Astronomy, Johns Hopkins University, 3400 N. Charles Street, Baltimore, MD 21218, USA}

%
\begin{abstract}
Magnetic reconnection is a process that changes magnetic field topology in
highly conducting fluids. Within the standard Sweet-Parker model, this process
would be too slow to explain observations (e.g. solar flares).  In reality, the
process must be ubiquitous as astrophysical fluids are magnetized and motions of
fluid elements necessarily entail crossing of magnetic frozen in field lines and
magnetic reconnection.  In the presence of turbulence, the reconnection is
independent of microscopic plasma properties, and may be much faster than
previously thought, as proposed in \cite{LazarianVishniac:1999} and tested in
\cite{Kowal_etal:2009, Kowal_etal:2012}. However, the considered turbulence in
the Lazarian-Vishniac model was imposed externally. In this work we consider
reconnection-driven magnetized turbulence in realistic three-dimensional
geometry initiated by stochastic noise. We demonstrate through numerical
simulations that the stochastic reconnection is able to self-generate turbulence
through interactions between the reconnection outflows. We analyze the
statistical properties of velocity fluctuations using power spectra and
anisotropy scaling in the local reference frame, which demonstrates that the
reconnection produces Kolmogorov-like turbulence, compatible with
\cite{GoldreichSridhar:1995} model. Anisotropy statistics are, however, strongly
affected by the dynamics of flows generated by reconnection process. Once the
broad turbulent region is formed, the typical anisotropy scaling $l_\parallel
\propto l_\perp^{2/3}$ is formed, especially for high resolution models, were
the broader range of scales is available. The decay of reconnection outflows to
turbulent-like fluctuations, characterized by different anisotropy scalings,
strongly depends on $\beta$ plasma parameter. Moreover, the estimated
reconnection rates are weakly dependent on the model resolution, suggesting that
no external processes are required to make reconnection fast.
\end{abstract}

%
\section{Introduction}
\label{sec:intro}

Magnetic reconnection is a key problem for the magnetohydrodynamic (MHD) theory.
It describes the interaction of magnetic flux tubes, and in particular it
describes what happens when magnetic flux tubes cross each other. It is
impossible to fully predict how magnetic fields evolve in turbulent
environments, ubiquitous in astrophysical fluids, without a solution to this
question.

The theory of magnetic reconnection can be traced back to the classical
Sweet-Parker scheme \citep{Parker:1957, Sweet:1958}, for which the magnetic
fluxes of opposite polarity are brought into contact over an extended region
parallel to the flux and characterized by the scale $\lambda$, while experience
diffusion and annihilation over the region of thickness $\delta$. This $\delta$
determines the scale of the mass outflow, that happens with the Alfv\'en speed
$V_A$. Using mass conservation one can estimate the speed of crossing flux as
$V_{rec, SP} \approx V_A \, \delta / \lambda$. With $\delta$ determined by
microphysics and $\lambda$ being an astrophysical scale it is evident that the
Sweet-Parker reconnection rate is negligibly small for most of astrophysical
conditions. Indeed, taking into account that $V_{rec, SP}$ is determined by
Ohmic diffusion, i.e. $V_{rec, SP} \approx \eta / \delta$, where $\eta$ is the
resistivity, one gets the classical Sweet-Parker rate $V_{rec, SP} \approx V_A
S^{1/2} \ll V_A$ with $S = L V_A / \eta$ being the Lundquist number, where $L$
is a scale of the flow. Given the highly nature of astrophysical plasmas,
Sweet-Parker reconnection predicts that reconnection occurs indeed very slowly.

For most of the astrophysical environments the Sweet-Parker reconnection
essentially means virtually no reconnection at finite timescales, which comes in
contradiction with the observations of e.g. solar flares, which the most natural
explanation is based on fast magnetic reconnection. Fast in this situation means
that magnetic reconnection rate $V_{rec}$ must not depend, or at least exhibits
very weak, e.g. logarithmic, dependence on the Lundquist number $S$.

Different suggestions have been made to solve the problem of fast reconnection.
They included solutions that involve making $\lambda$ small, i.e. similar to
$\delta$ by bending magnetic field towards the reconnection point at a sharp
angle, as an ingenious suggestion by \cite{Petschek:1964}. The corresponding
magnetic configuration was termed ``X-point reconnection''. A list of other
solutions where the magnetic reconnection can achieve large speeds due to a
particular configuration of magnetic fluxes can be found in the book by
\cite{PriestForbes:2007}.

The main limitation of these models is that they deal with the reconnection in
rather special environments and do not address the problem of magnetic
reconnection in generic astrophysical/turbulent circumstances. Could turbulence
make reconnection fast? This issue was discussed in a number of papers. While
\cite{JacobsonMoses:1984} dealt with the effects of turbulence on Ohmic
resistivity, therefore decreasing $S$ by some factor,
\cite{MatthaeusLamkin:1985, MatthaeusLamkin:1986} performed 2D numerical
simulations of MHD turbulence and claimed that the formation of X-points in 2D
made the reconnection fast. Due to significant differences of MHD turbulence
nature in 2D and 3D, as well as the ambiguity of reconnection rate measurements
within the 2D distribution of turbulent magnetic flux, these interesting
approaches could not be directly compared to real astrophysical scenarios.

A following model that related the modern theory of 3D MHD turbulence, i.e. the
\cite{GoldreichSridhar:1995} one, and magnetic reconnection was proposed in
\citet[][henceforth LV99]{LazarianVishniac:1999}. There the outflow region
$\delta$ is not determined by microscopic diffusive processes, but by the
wandering of magnetic field lines. The prediction of the LV99 theory was that
the reconnection changes with the level of turbulence. This quantitative
prediction was successfully tested in the numerical studies
\cite{Kowal_etal:2009, Kowal_etal:2012} investigating the effects of resolution,
explicit resistivity, ways of turbulence is driven, and strengths of guide
field. More recently, another confirmation of LV99 came from relativistic MHD
simulations by \cite{Takamoto_etal:2015}. Observational testing of the theory
are discussed e.g. in recent reviews by \cite{Lazarian_etal:2015,
Lazarian_etal:2016}.

The LV99 predicts that reconnection is fast, i.e. does not depend on $S$, in
generic astrophysical conditions. It predicts that reconnection happens not only
at particular places where the magnetic field lines happen to undergo a special
configuration, but through the entire turbulent volume. In fact, the theory
makes the MHD turbulence theory by \cite{GoldreichSridhar:1995} self-consistent
and predicts the violation of the classical magnetic flux freezing
\cite{Alfven:1942}\footnote{The violation of flux freezing is implicit in LV99
but is explicitly treated in the subsequent proceedings paper
\cite{VishniacLazarian:1999}.}. The violation of flux freezing in turbulent
fluids entails many vital astrophysical consequences \cite[see][]{Lazarian:2005,
Santos-Lima_etal:2010, Lazarian_etal:2012} and it has been explored
theoretically \cite{Eyink:2011, Eyink_etal:2011} and confirmed numerically
\cite{Eyink_etal:2013}.

While turbulence is really ubiquitous in astrophysical environments
\cite[see][]{Armstrong_etal:1995, Padoan_etal:2009, ChepurnovLazarian:2010,
Chepurnov_etal:2015}, a question arises of whether reconnection itself could
induce turbulence that would make the it fast. This possibility was mentioned in
LV99 and quantified in \cite{LazarianVishniac:2009} within the model of flares
of magnetic reconnection in high $\beta$-plasma environments\footnote{Plasma
$\beta$ is the ratio of the thermal to magnetic pressure.}. The first numerical
simulations of magnetic reconnection induced by turbulence that is generated by
reconnection were performed in \cite{Beresnyak:2013} under incompressible
approximation, and in \cite{Oishi_etal:2015} and \cite{HuangBhattacharjee:2016}
using compressible codes. However, these works did not consider a few studies
which we address here. For instance, \cite{Beresnyak:2013} \&
\cite{Oishi_etal:2015} did not perform the studies of turbulence properties and
anisotropy statistics. Although \cite{HuangBhattacharjee:2016} included these
studies, their setup was significantly different, and the statistics were
calculated at relatively earlier time $t = 3.5 t_A$. Moreover, none of these
works addressed the question of turbulence development under different
$\beta$-plasma parameters.

In this paper we aim to understand the reconnection-driven turbulence in
compressible MHD framework, focusing on long time evolution and turbulence
statistics changes during the whole simulation. For that we present a number of
high resolution numerical experiments, for which we analyze and interpret the
properties of obtained turbulent velocity fluctuations. We demonstrate the
dependencies on the resolution and $\beta$-plasma. In Section~\ref{sec:model} we
present our methodology and preformed numerical models. In
Section~\ref{sec:results} we describe obtained results from the analysis of
temporal evolution and statistical properties of generated turbulent
fluctuations. In Section~\ref{sec:discussion} we discuss our results and compare
them to the previously done models. Finally, in Section~\ref{sec:conclusions} we
draw our conclusions.

%
\section{Methodology and Modeling}
\label{sec:model}

We use a high-order shock-capturing Godunov-type code AMUN\footnote{The code is
freely available at \href{http://amuncode.org}{http://amuncode.org}} based on
the adaptive mesh. The code integrates the set of isothermal compressible
magnetohydrodynamic (MHD) equations
\begin{eqnarray}
 \pder{\rho}{t} + \nabla \cdot \left( \rho \vc{v} \right) = 0,
 \label{eq:mass} \\
 \pder{\rho \vc{v}}{t} + \nabla \cdot \left[ \rho \vc{v} \vc{v} +
 \left( a^2 \rho + \frac{B^2}{8 \pi} \right) I - \frac{1}{4 \pi} \vc{B} \vc{B}
 \right] =  \label{eq:momentum} \\ \nu \nabla^2 \left( \rho \vec{v} \right), \nonumber \\
 \pder{\vc{B}}{t} + \nabla \times \vc{E} = 0, \label{eq:induction}
\end{eqnarray}
where $\rho$ and $\vc{v}$ are plasma density and velocity, respectively,
$\vc{B}$ is the magnetic field, $\vc{E} = - \vc{v} \times \vc{B} + \eta \,
\vc{J}$ is the electric field, $\vc{J} = \nabla \times \vc{B}$ is the current
density, $a$ is the isothermal speed of sound, and $\nu$ and $\eta$ are the
viscosity and resistivity coefficients, respectively.

We integrated the governing equations using the $3^{rd}$ order 4-stage Strong
Stability Preserving Runge-Kutta (SSPRK) method \citep{Ruuth:2006}, a 3D spatial
Gaussian processes based reconstruction limited near extrema \citep{Kowal:2016},
and a multi-state Harten-Lax-van Leer (HLLD) approximate Riemann solver
\cite{Mignone:2007}.  In order to keep $\nabla \cdot\vec{B}$ negligible, we
solve the induction equation (Eq.~\ref{eq:induction}) using the hyperbolic
divergence cleaning based on generalized Lagrange multiplier (GLM) method
\citep{Dedner_etal:2002}. This set of numerical algorithms, together with the
stability CFL-coefficient set to $0.3$ for all models, results in stable
numerically modeled data of high quality.

We performed a number of numerical simulations within a 3D domain with physical
dimensions $1.0 \times 4.0 \times 0.5$ with its center placed at Cartesian
coordinates $(0, 0, 0)$. Our code make use of adaptive mesh, for which the
refinement criterion is based on the magnitude of the local vorticity
$|\vec{\omega}| = |\nabla \times \vec{V}|$ and current density $|\vec{J}|$. The
thresholds determining the refinement and derefinement of the mesh are set to
$0.1$ and $0.01$, respectively, indicating that the mesh is quickly refined in
the turbulent and reconnection regions. The base domain is divided into $2
\times 8 \times 1$ blocks, each with $32^3$ resolution, from which we refine the
mesh using the above mentioned criterion up to 3, 4, and 5 refinement levels
with the resulting effective grid sizes $\Delta x = \Delta y = \Delta z = h =
1/256$, $1/512$, and $1/1024$, respectively. For the later, the effective
resolution is $1024 \times 4096 \times 512$, therefore. For models with
different $\beta$-plasma parameters, indicated with $\diamond$ in the figures,
we use box which is squared in the XZ-plane, i.e. its physical dimensions are
$1.0 \times 4.0 \times 1.0$, with the maximum refinement level equal to 4, and
the effective resolution $512 \times 2048 \times 512$.

The initial magnetic field configuration is antiparallel along the X direction
with the magnitude equal to $1.0$ and a discontinuity placed at the XZ plane
$y=0$. Additional guide field is imposed along the Z direction with a uniform
amplitude $0.1$. Uniform initial density is set to $1.0$ in the whole
computational domain. The initial velocity perturbation with random distribution
of directions and maximum amplitude of $0.0173$\footnote{Each component of
velocity has a uniform random distribution between $-0.01$ and $0.01$ resulting
in a non-uniform velocity magnitude distribution with a maximum value of $0.01
\sqrt{3}$.} is set in the region within the distance of $0.1$ from the initial
magnetic field discontinuity. We set the sound speed $a$ to $0.5$, $1.0$ or $4.0$,
which give plasma parameter $\beta = p_{\rm th} / p_{\rm mag} \approx 0.5$,
$2.0$ or $32.0$, respectively.

We do not set viscosity and resistivity explicitly. The dissipation of kinetic
and magnetic energies happens through the numerical diffusion. From earlier
models \cite[see][]{Kowal_etal:2009, Kowal_etal:2012} we estimated the numerical
resistivity to be smaller than $3\cdot10^{-4}$ in models with the grid size $h =
1/256$, resulting in Lundquist number $S \approx 3 \cdot 10^{3}$ or higher for
models with the grid sizes presented here. Since the same scheme was applied in
the solution of all MHD equations, we can assume that the numerical viscosity
and resistivity are similar, which gives the Prandtl number $Pr_m = \nu / \eta$
of the order of one.

\begin{figure}[t]
\centering
\includegraphics[width=0.48\textwidth]{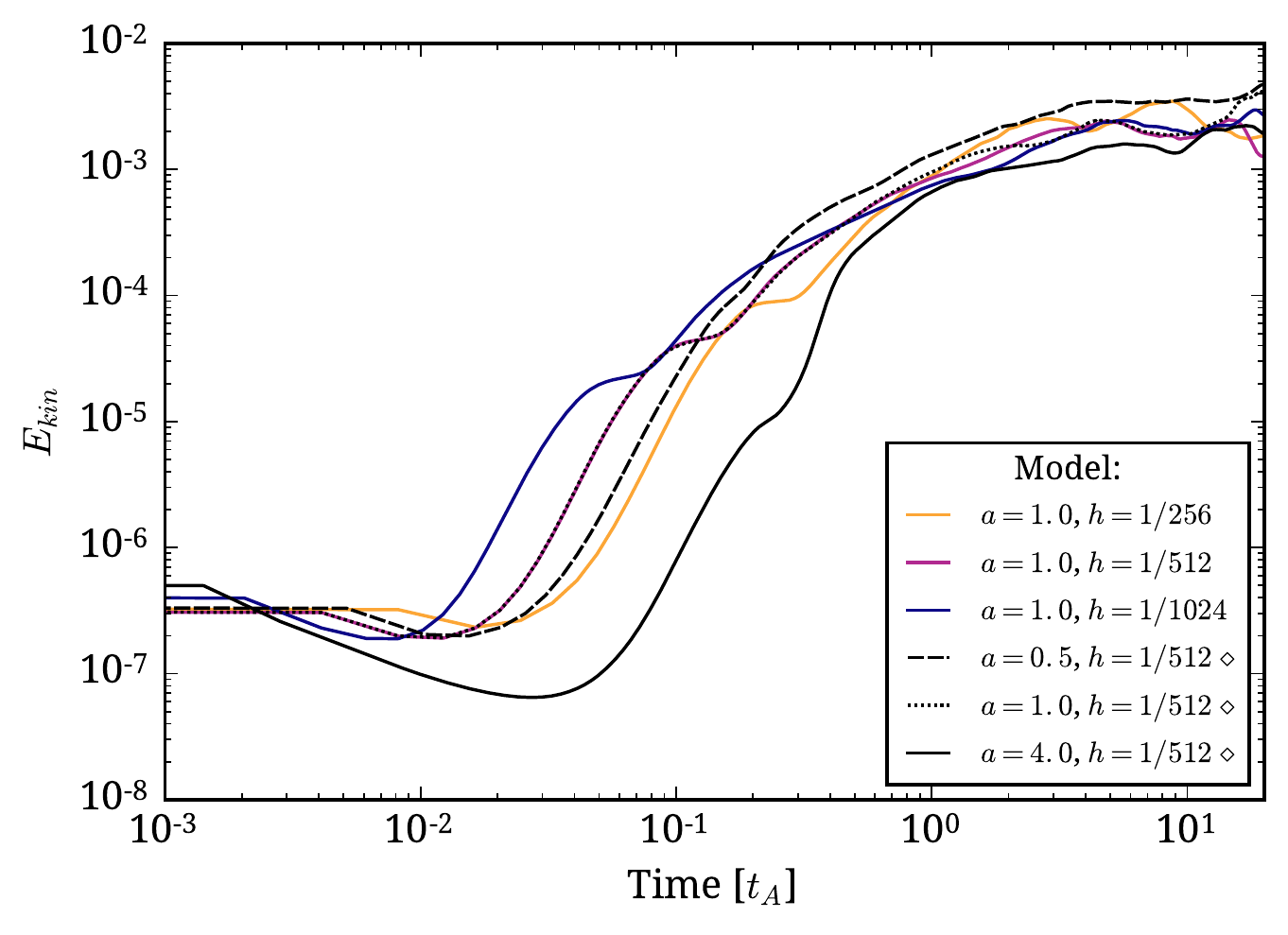}
\caption{Evolution of kinetic energy density for models with different
resolutions and sound speeds. Depending on the grid size $h$ or $\beta$ plasma
parameter (determined by sound speed $a$) the growth of kinetic energy starts at
different times. In all cases, the kinetic energy grows from the initial value
of around $5\cdot10^{-7}$ by more than three orders of magnitude, independently
of $h$ or $\beta$. Models with different sound speeds $a$ are indicated by
symbol $\diamond$. They are also performed in a box with $L_x = L_z = 1.0$.
\label{fig:ekin_evol}}
\end{figure}

We applied periodic boundary conditions along the X and Z directions and open
(normal derivatives to the boundary plane are set to zero) ones along the Y
direction. This choice of boundary conditions corresponds to a subvolume of a
larger box with a size $L \gg 1.0$ and imposes certain restrictions on
applications, which we discuss more extensively in the discussion section.

%
\section{Analysis and Results}
\label{sec:results}

%
\subsection{Turbulence Evolution}

We started our models with a weak random velocity fluctuations of the order of
1\% of the Alfv\'en velocity, which launches the magnetic reconnection process
within the initial current sheet.  This process normally converts magnetic
energy into kinetic one and heat. Since we used the isothermal equation of
state, the set up corresponds to a situation that the heat is quickly
dissipated.

The first question we want to answer is, if the process of magnetic reconnection
can sustain the injection of kinetic energy to the system or if the initial
velocity fluctuations simply decay due to the dissipation. In order to answer
it, we analyze our result first in terms of global quantities, such as the
kinetic energy evolution, the evolution of velocity component contribution to
the kinetic energy, the evolution of total vorticity generated by the
reconnection, and the change of the volume fraction characterized by a high
vorticity value.

\begin{figure}[t]
\centering
\includegraphics[width=0.48\textwidth]{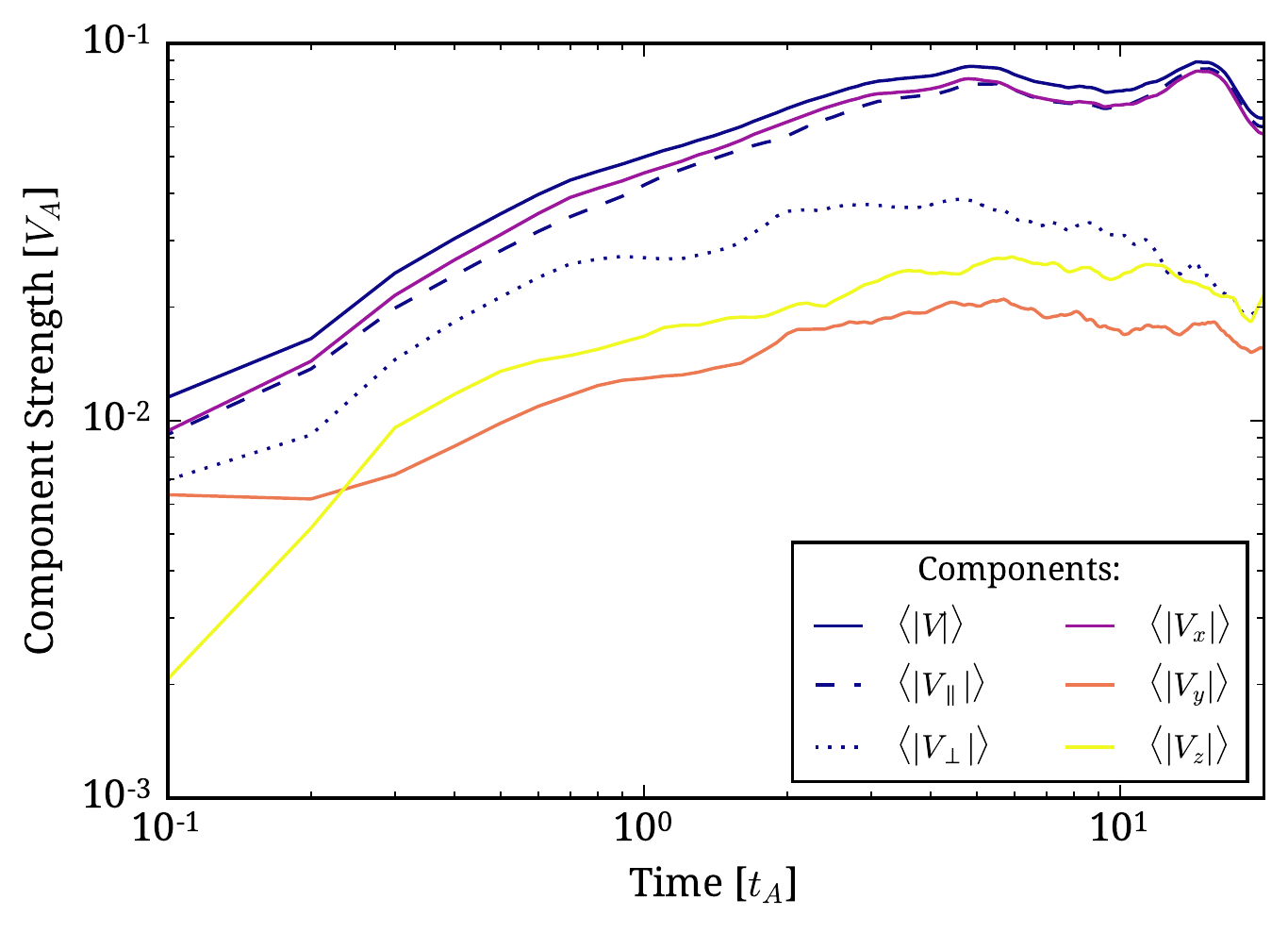}
\caption{Evolution of standard deviation of the velocity and its components: X,
Y, Z, and parallel and perpendicular to the local magnetic field for model with
sound speed $a = 1.0$ and the grid size $h = 1/512$. \label{fig:velo_comp}}
\end{figure}

Figure~\ref{fig:ekin_evol} shows the evolution of kinetic energy density
integrated over the whole domain, up to the final simulation time $t=20.0$. The
initial value of the energy for all models is about $5\cdot10^{-7}$. We shall
remind, that the turbulence is generated only in the vicinity of current sheet
by the action of magnetic reconnection. Therefore, it can be dynamically
important there. In most of the domain volume, however, the kinetic energy
remains negligible. The figure demonstrates a quick growth of $E_{kin}$ to
values above $10^{-3}$ at relatively short time of one Alfv\'en time unit. This
indicates an increase of the kinetic energy by almost three orders of magnitude.
We should note, that the initial velocity perturbations were generated from a
random noise, therefore they are strongly dissipated by the numerical
dissipation during an initial period of about $0.02 t_A$, causing a small drop
of kinetic energy, and reducing the strength of fluctuations interacting with
the current density, as seen in Figure~\ref{fig:ekin_evol}. This drop depends on
the used grid size $h$ and sound speed $a$. For example, it is bigger and
reaches the minimum at slightly later times (the minimum appears around $0.03
t_A$) for model with the sound speed equal $4.0$ (black line), comparing to
model with $a=1.0$. Even though these fluctuations are weakened, they are still
strong enough to sufficiently deform the current sheet locally, which results in
development of instabilities. For the scope of this manuscript, we do not
identify the type of instability, and just focus on the characteristics
turbulence produced as a result. After around $1 t_{A}$, the levels of kinetic
energy are comparable for models with different sound speeds $a$ (or $\beta$)
and grid sizes $h$.

\begin{figure*}[t]
\centering
\includegraphics[width=0.48\textwidth]{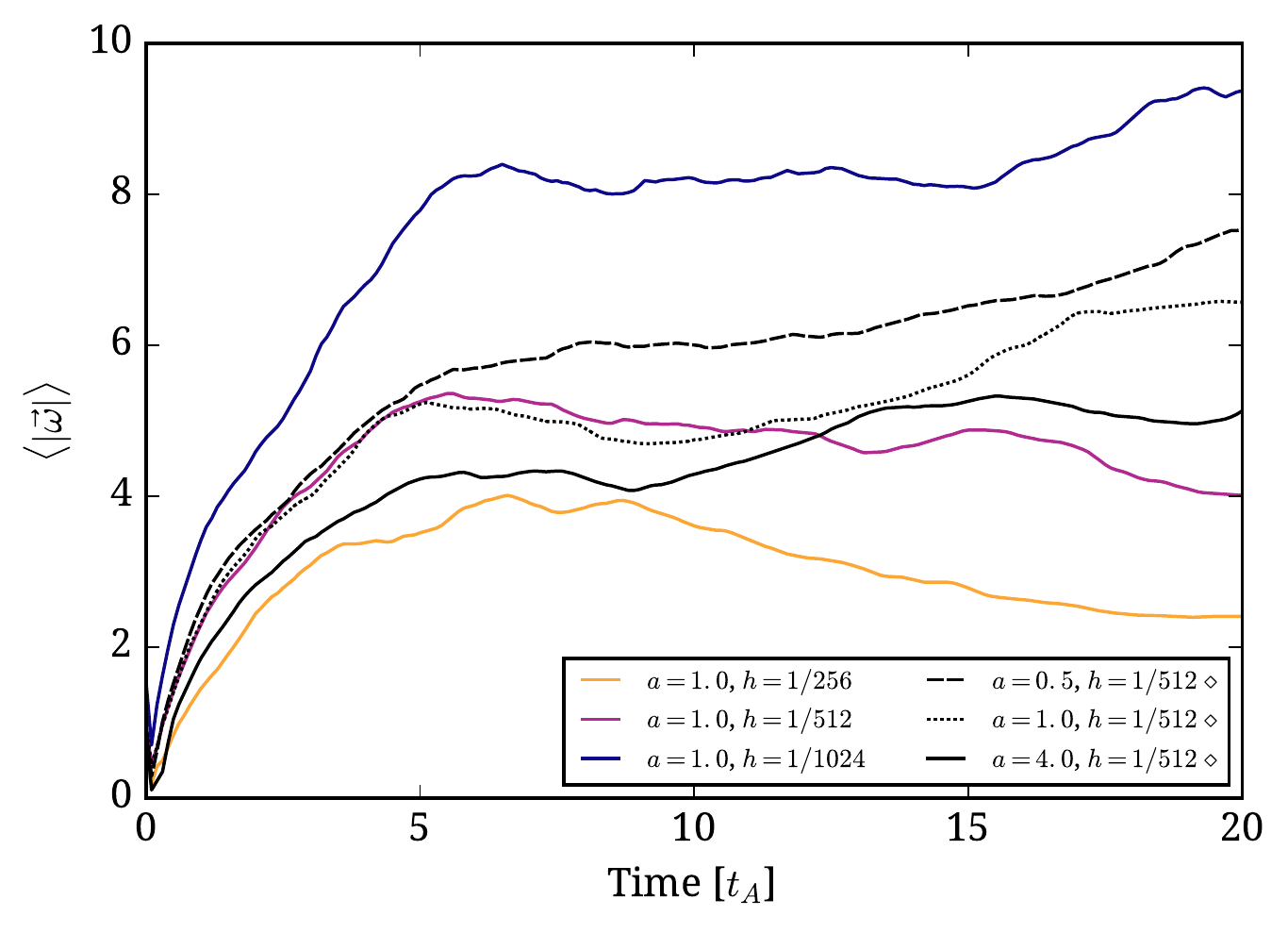}
\includegraphics[width=0.48\textwidth]{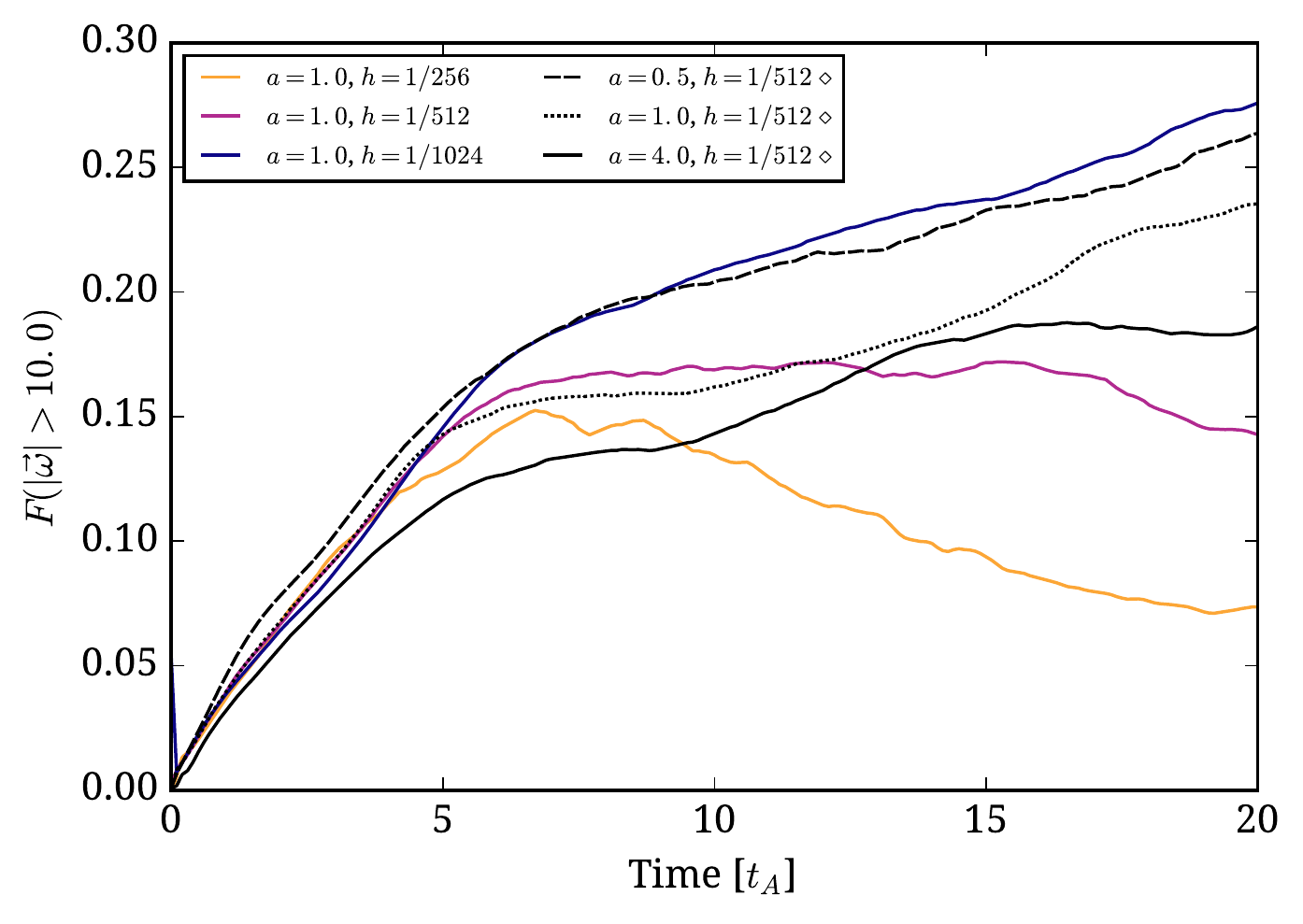}
\includegraphics[width=0.48\textwidth]{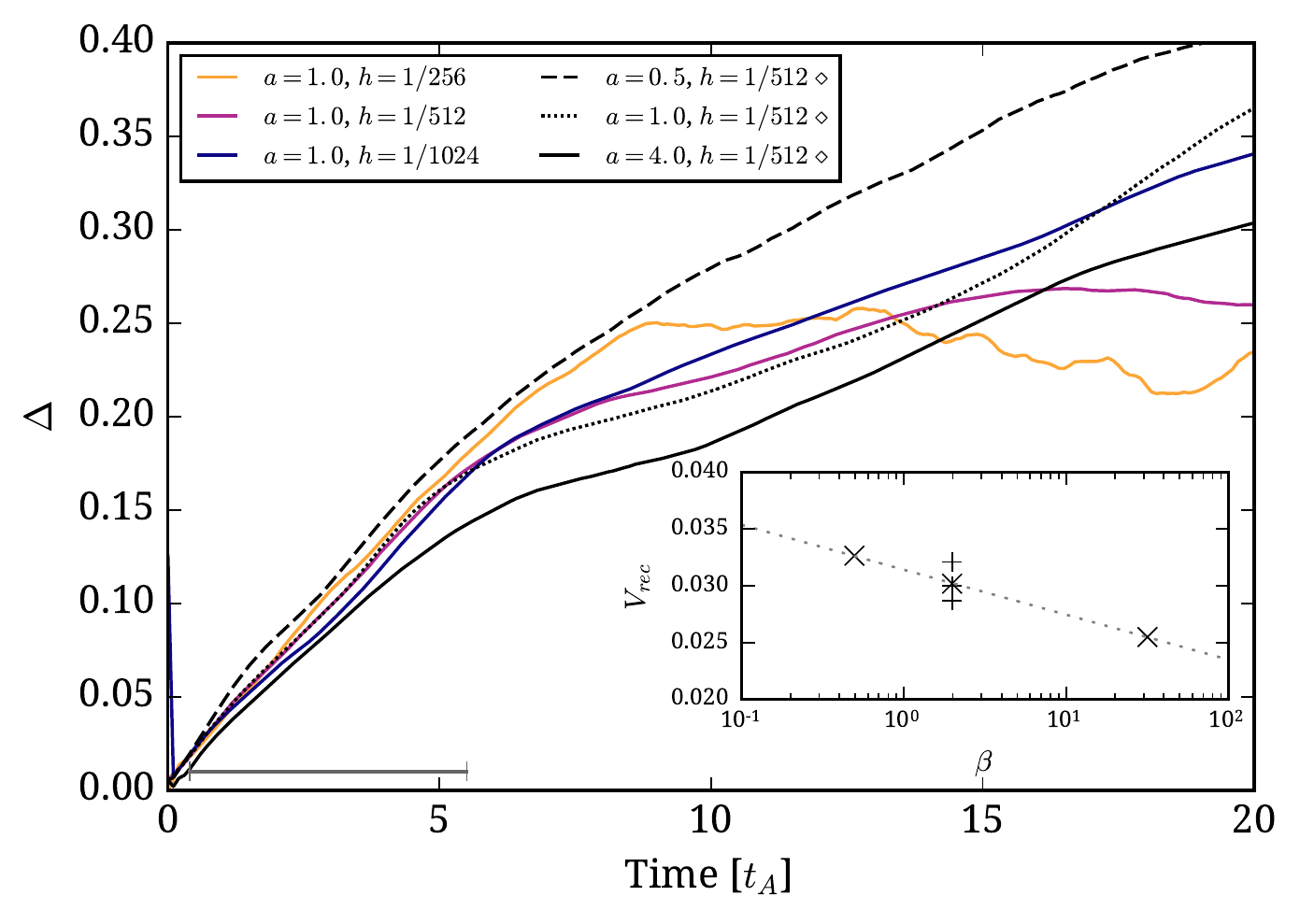}
\includegraphics[width=0.48\textwidth]{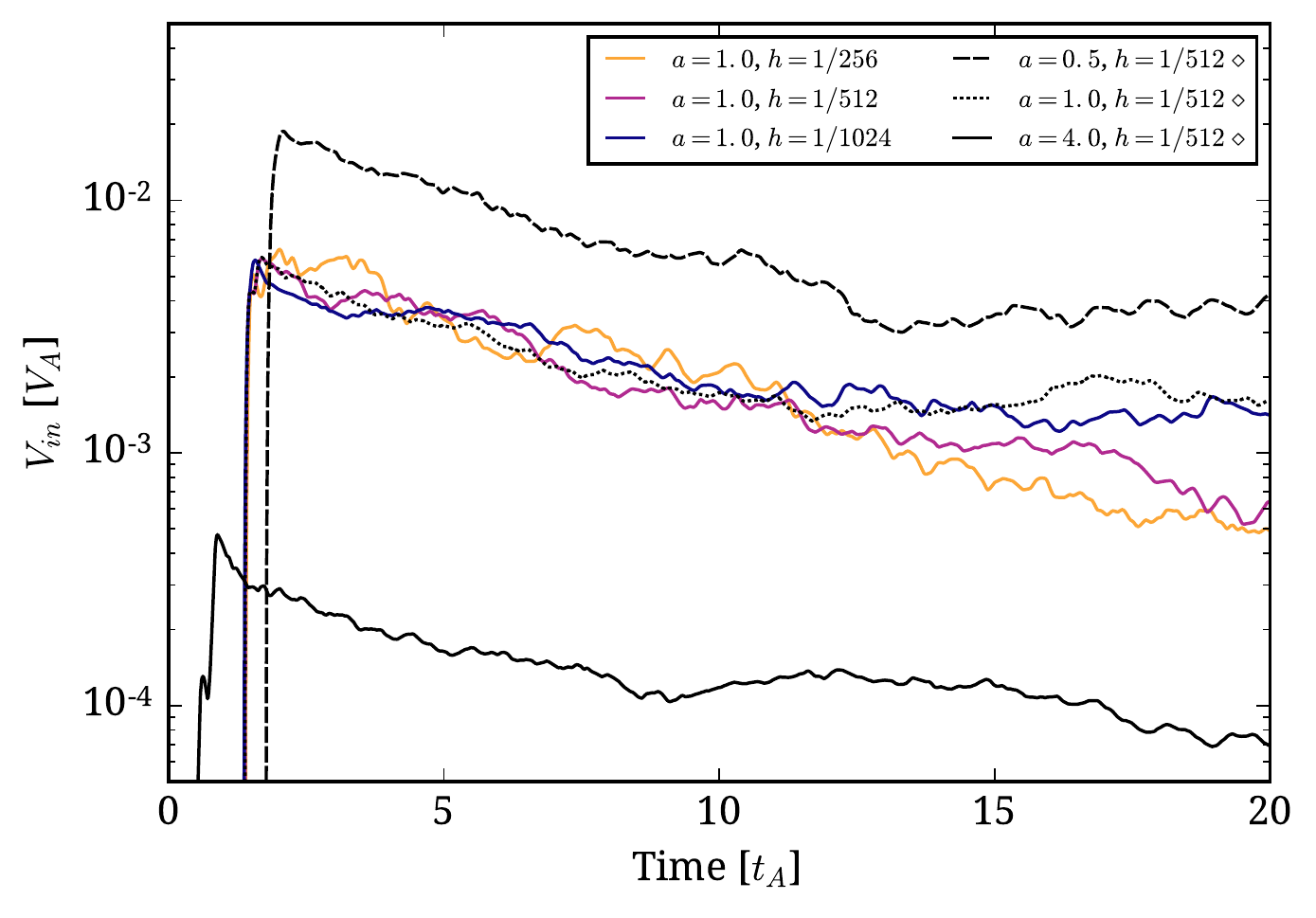}
\caption{Evolution of the mean vorticity for models with different resolutions
(top left) and the filling factor of volume where the vorticity is equal or
larger than a given threshold (top right). The plot shows the absolute value of
vorticity averaged over the simulation box for models with different
resolutions. In the bottom left plot we show the evolution of the turbulent
region thickness $\Delta$ for models with different resolutions. In the subplot
of this plot we estimated the reconnection rate measured by the growth rate of
the turbulent region thickness $V_{rec} = d \Delta / dt$, in a similar way as
\cite{Beresnyak:2013}. The estimation was done within the period indicated by
the horizontal gray line. In the bottom right plot we show the average inflow
speed $V_{in}$, i.e. the speed with which the new magnetic flux is brought into
the box through the top and bottom open boundaries. This speed is expressed in
$V_A$ and is useful to estimate the magnetic field dissipation rate. As we can
see, $V_{in}$ seems to be independent of the grid size (or numerical
dissipation), but is strongly dependent on sound speed $a$.
\label{fig:vort_evol}}
\end{figure*}

How this energy is distributed among different velocity components? In
Figure~\ref{fig:velo_comp} we show the standard deviation of the total velocity,
its X, Y, and Z components, and components parallel and perpendicular to the
local field for model with sound speed $a = 1.0$ and the grid size $h = 1/512$.
Initially, the X, Y, and Z-components of velocity are equal, but shortly the X
component, which is in the direction of the antiparallel magnetic field $B_x$,
starts to dominate. Comparing to the kinetic energy evolution, we see that this
component caries most of the energy. The Z-component is slightly stronger than
the Y one, since we imposed non-zero guide field in the system. Comparing
components parallel (dashed) and perpendicular (dotted) to the local field, we
see that the parallel one is developing to values a few times larger than the
perpendicular one. This indicates that the strongest motion generation takes
place along the mean magnetic field and is associated with the reconnection
process. A similar behavior is observed for two components, $V_{x}$ and
$V_\parallel$.

The evolution of kinetic energy and velocity components described above clearly
demonstrates that the reconnection is able to generate significant plasma
motions, which could dominate the local dynamics. Below, we investigate more
closely what type of motions we observe in our models, turbulent or laminar. The
simplest way of detecting turbulent eddies in a system is by calculating the
vorticity, $\vec{\omega} = \nabla \times \vec{v}$, which directly measures the
rotational motions. In the first analysis, we calculate the magnitude of
vorticity and average it over the computation domain at each time step. We also
analyze the reconnection rate estimated using two different methods. Later, we
analyze their spectral properties and anisotropy.

The top left plot of Figure~\ref{fig:vort_evol} shows the average of vorticity
magnitude $|\vec{\omega}|$ as a function of time for all models. The amount
of turbulent motions generated due to reconnection is growing until it reaches a
saturation at around $t=5.0$. The vorticity growth rate seems to gradually
decrease with time until the saturation. Since the reconnection works at
smallest scales, the vorticity growth depends on the resolution, what can be
seen by different maximum values of averaged vorticity, $|\vec{\omega}| \approx
4.0$, $5.0$, and $8.0$ for models with grid sizes $h = 1/256$, $1/512$, and
$1/1024$, respectively. This dependence with the grid size is not linear and may
indicate that for smaller grid sizes even higher vorticities can be generated.
After reaching the maximum, the vorticity generation slowly decays for models
with medium and large grid sizes. It increases for the model with the highest
resolution, however. In three models with different sound speeds (dashed,
dotted, and solid black lines corresponding to $a = 0.5$, $1.0$, and $4.0$,
respectively), the total vorticity grows through the whole simulation, but its
value is $a$ dependent, i.e. it decays with $a$ (or $\beta$).

\begin{figure*}[t]
\centering
\includegraphics[width=0.96\textwidth]{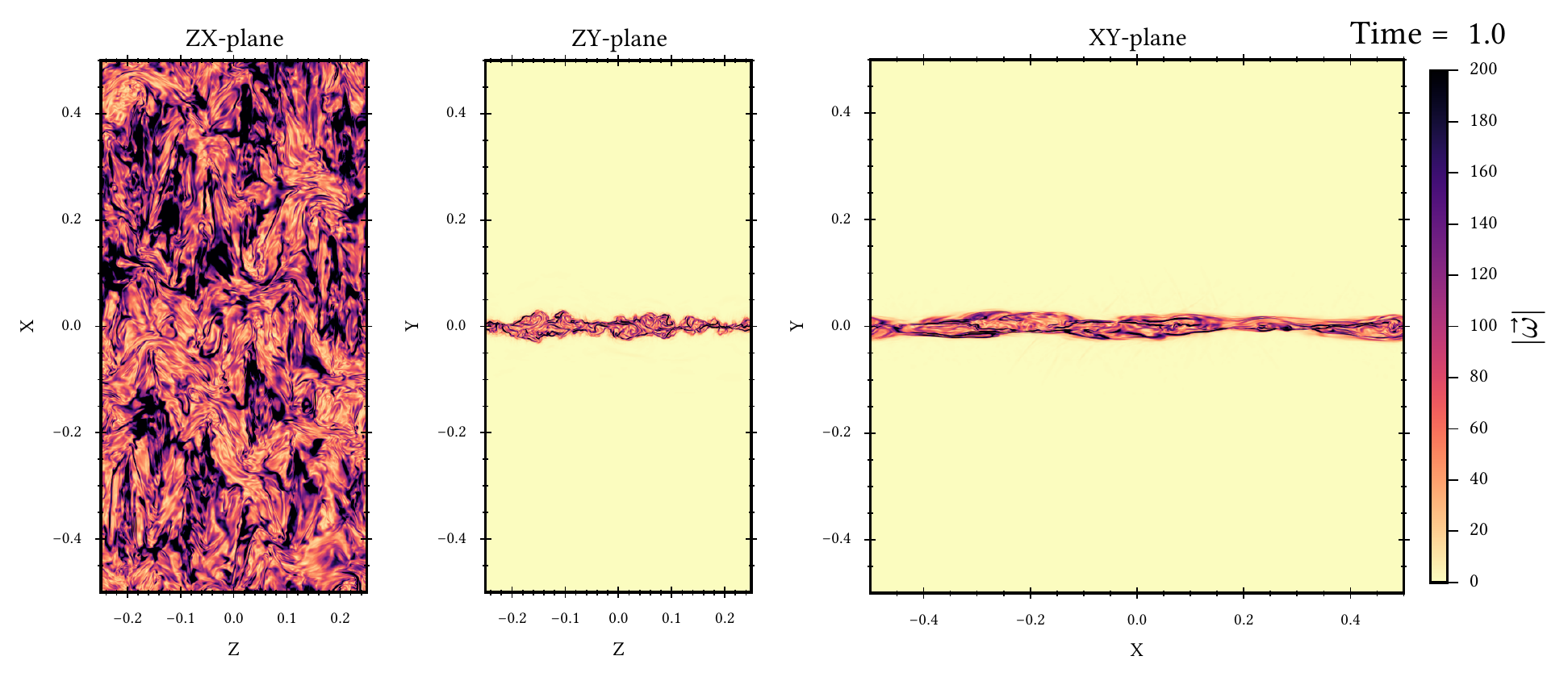}
\includegraphics[width=0.96\textwidth]{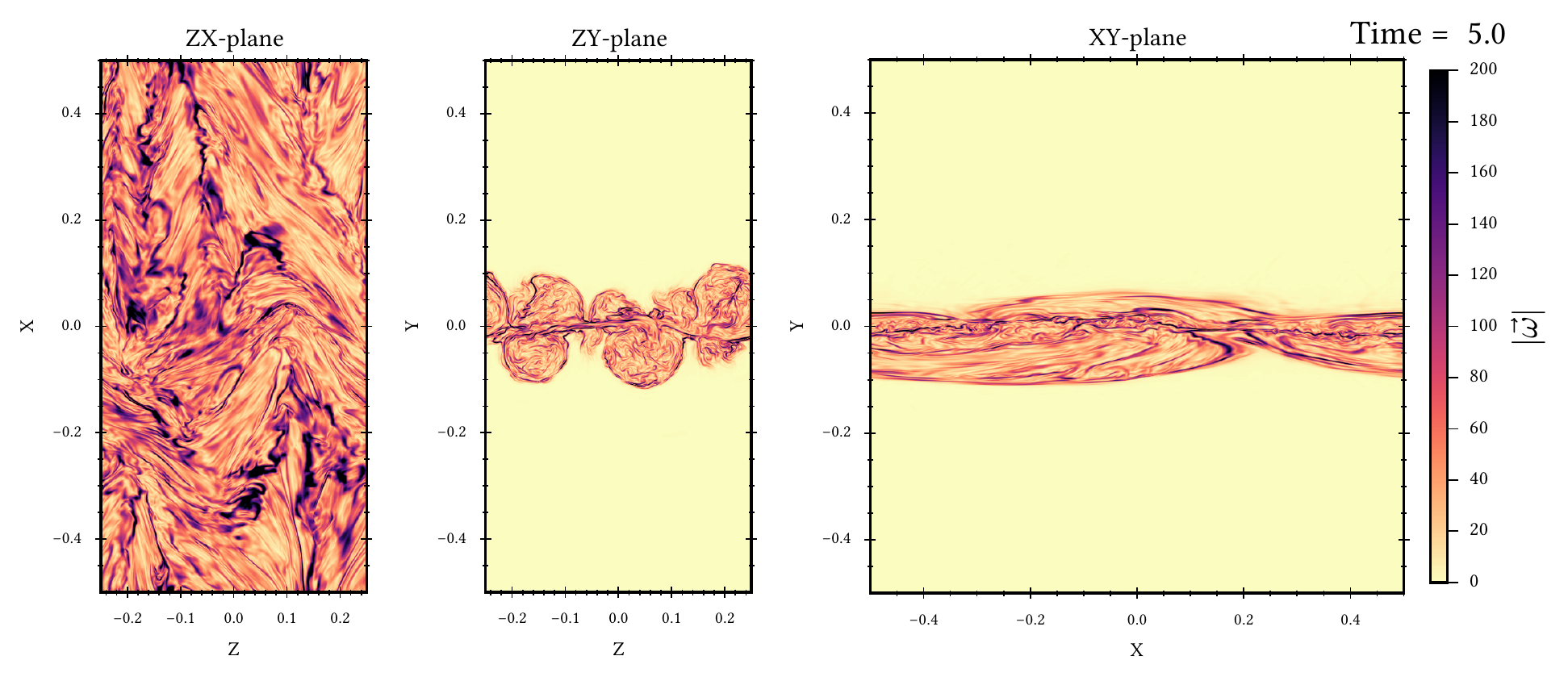}
\includegraphics[width=0.96\textwidth]{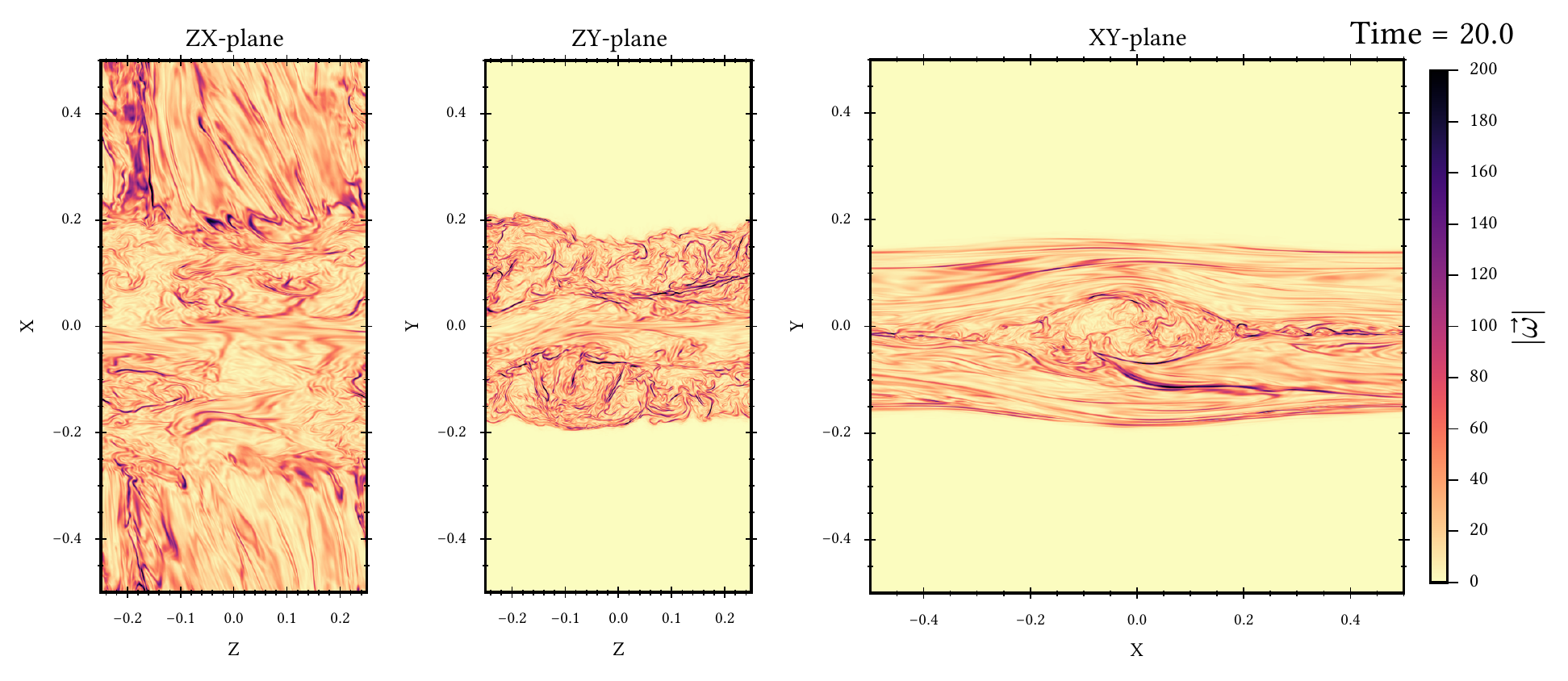}
\caption{Vorticity amplitude slices along three main planes, ZX (left), ZY
(center), and XY (right) for three different moments in time, $t = 1.0$, $5.0$,
and $20.0$ (top, middle, and bottom, respectively) for model with sound speed $a
= 1.0$ and grid size $h = 1/1024$. The turbulent region develops near the
initial current sheet and expands in the Y direction. The plots show subregions
limited to $y \le 0.5$ from the whole domain extended up to $y = 2.0$.
\label{fig:vorticity_visualization}}
\end{figure*}

The top right plot of the same figure shows the filling factor for vorticity
$|\vec{\omega}| \ge 10.0$. It simply shows a volume where vorticity is equal or
larger than the threshold, and indicates how the volume of strong rotational
motions changes. The chosen threshold can be interpreted as a change of velocity
by $0.1 V_{A}$ within a distance $0.01$, which corresponds to relatively strong
rotation or shear. In our simulations, we observe peak values of vorticity as
high as a few hundreds, related to very strong reconnection events (change of
the order of $V_A$ at the grid size scale). The top right plot in
Figure~\ref{fig:vort_evol} demonstrates the initial nearly linear growth of the
volume occupied by relatively strong rotational motions, almost independent of
the used resolution or sound speed, until about $t = 6.0$. This moment, at which
the growth rate of the filling factor suddenly changes, is also independent on
the effective resolution $h$. The maximum value, however, is resolution
sensitive, with the highest volume observed in the model with the smallest grid
size. The evolution of filling factor after this moment have different behavior
for different resolution, as well. For the model with the largest grid size, the
filling factor starts to decay after $t \approx 8$, for middle grid size it stay
nearly constant until $t \approx 16$, and for the smaller grid size, it seems to
continue growing. This indicates that in order to resolve well the turbulence we
have to use the highest resolution possible. The filling factor evolution seems
to be very weakly dependent on plasma $\beta$ parameter. Comparing models with
different $a$ and the same grid size $h = 1/512$, we see that the filling factor
follows similar evolution growing in all cases, but reaching different filling
factor values for different sound speeds. Comparing to the model with $a = 0.5$
(dashed line) to models with $a = 1.0$ and $a = 4.0$, the filling factor reaches
values by 5--20\% and 20--30\% lower, respectively, after $t = 5.0$.

In two bottom plots of Figure~\ref{fig:vort_evol} we show two ways of estimating
the reconnection rate. In the right plot of Figure~\ref{fig:vort_evol} we
present the evolution of the estimated thickness of the turbulent layer
$\Delta$. The thickness was calculated from the region where the magnitude of
vorticity is above a threshold, in our case $|\vec{\omega}| \ge 10.0$. As we
see, initially the thickness continues to grow linearly, almost independently of
the grid size, and after $t \approx 6.0$ its growth slows down. For the lowest
resolution model ($h = 1/256$), it continues to grow up to $t = 9.0$ and then
saturates. In the case of medium resolution model ($h = 1/512$), $\Delta$
continues to increase until $t \approx 16$, and for the highest resolution model
($h = 1/1024$) it increases with the same rate until the end of simulation. For
models with different sound speeds, the turbulent region thickness grows during
the whole simulation. In the subplot of this plot, we show reconnection rates
for all models estimated from the growth rate of the turbulent region thickness
$V_{rec} = d \Delta / dt$ within the initial period $t \le 5.0$, in the same way
as it was done by \cite{Beresnyak:2013}. The $\times$ points correspond to
models with different $\beta$, while the $+$ points are for models with
different resolutions and the same $a = 1.0$. We see that the reconnection rate
depends slightly on $\beta$ plasma, decreasing with its value. We estimated
$V_{rec}$ to be about $0.0327\pm0.0003$, $0.0302\pm0.0002$, and
$0.0255\pm0.0003$ for $\beta \approx 0.5$, $2.0$, and $32.0$, respectively. From
these points we estimated an empritical dependence $V_{rec}(\beta) = -0.0017
\log \beta + 0.0314$, with both coefficients rounded to the fitting error. This
weak dependence could indicate the importance of supersonic motions on bending
magnetic field lines within the current sheet. There the reconnection outflows
can approach the Alfv\'en speed, which means that for the low $\beta$ plasma,
the outflows become supersonic. Also, the magnetic field strength is reduced,
therefore its easier to bend and effectively reconnect its lines.

The reconnection rate $V_{rec}$ also weakly depends on the used resolution, i.e.
it grows with the grid size $h$. The resolution dependence could be related to
the numerical resistivity, which is proportional to $h$, causing the
reconnection efficiency to be higher for models with bigger $h$. For the model
with the highest resolution ($h = 1/1024$), the estimated reconnection rate is
about $0.0287\pm0.0003$, nearly twice as much as the value estimated by
\cite{Beresnyak:2013}. Remembering that \cite{Beresnyak:2013} used much less
dissipative spectral code and smaller values of the resistivity, our result
should be compatible with theirs.

As we indicated, the reconnection rates presented in the bottom left plot were
estimated for times $t \le 5.0$. At later times, the growth rate of turbulent
regime depends on the grid size $h$. Once the turbulence is fully developed, we
should expect the reconnection to perform more efficiently, as predicted by
LV99. The observed tendency is that for $t \ge 5.0$ the reconnection rate
estimated using the thickness of turbulent region decreases with $h$, and even
statures in the case of the lowest resolution model ($h = 1 / 256$). For the
highest resolution model, the turbulent region thickness $\Delta$ increases all
the time, yet its growth rate is not as high as at earlier period $t \le 5.0$.
In real systems, however, where the resolution is virtually infinite, we should
expect that the thickness of turbulent region grows constantly with the same
rate for later times too. In models with different sound speeds we also see
decrease of the growth rate of the turbulent region thickness after $t \approx
5.0$, however, it is roughly independent of the used $\beta$.

\begin{figure*}
\centering
\includegraphics[width=0.48\textwidth]{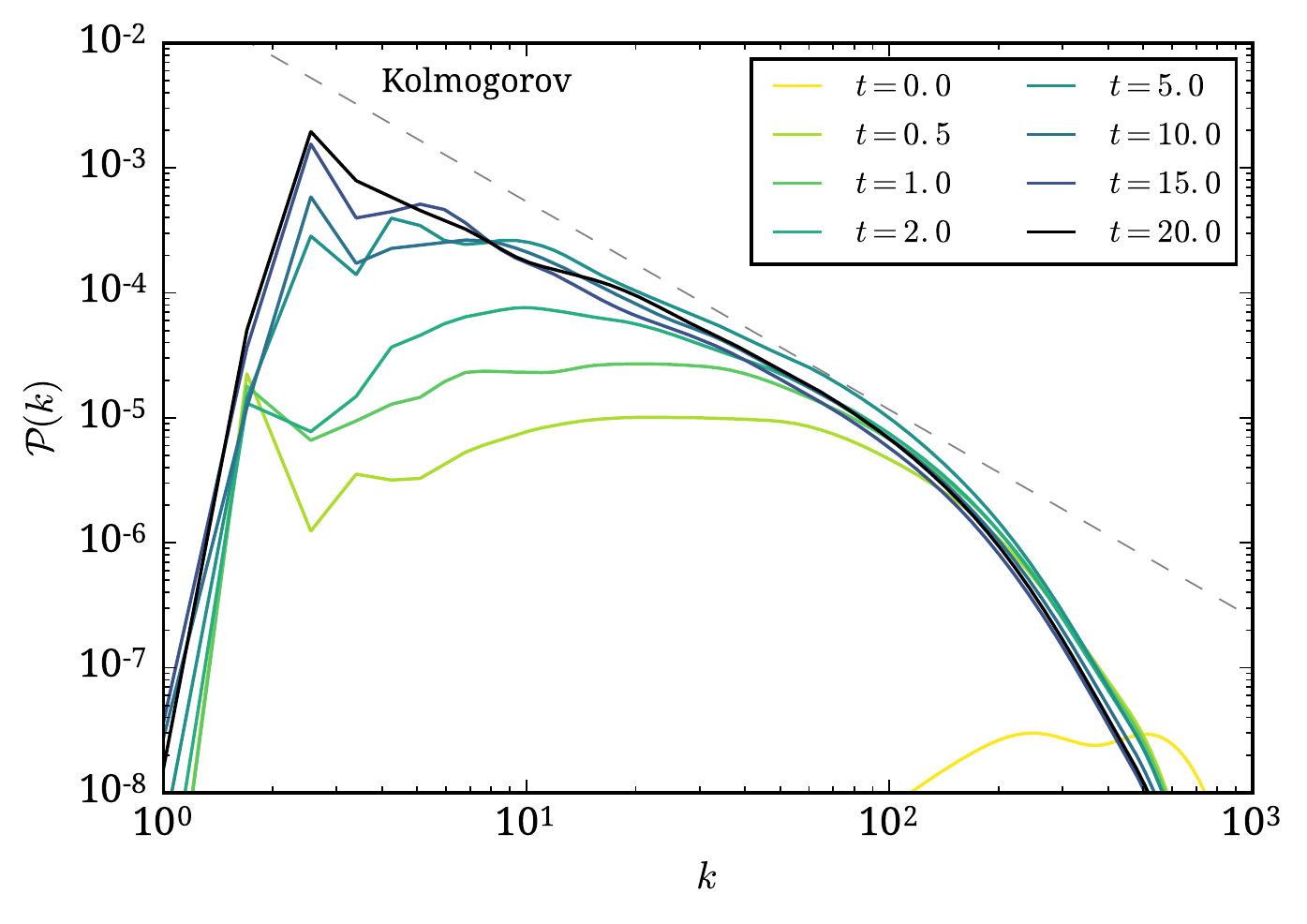}
\includegraphics[width=0.48\textwidth]{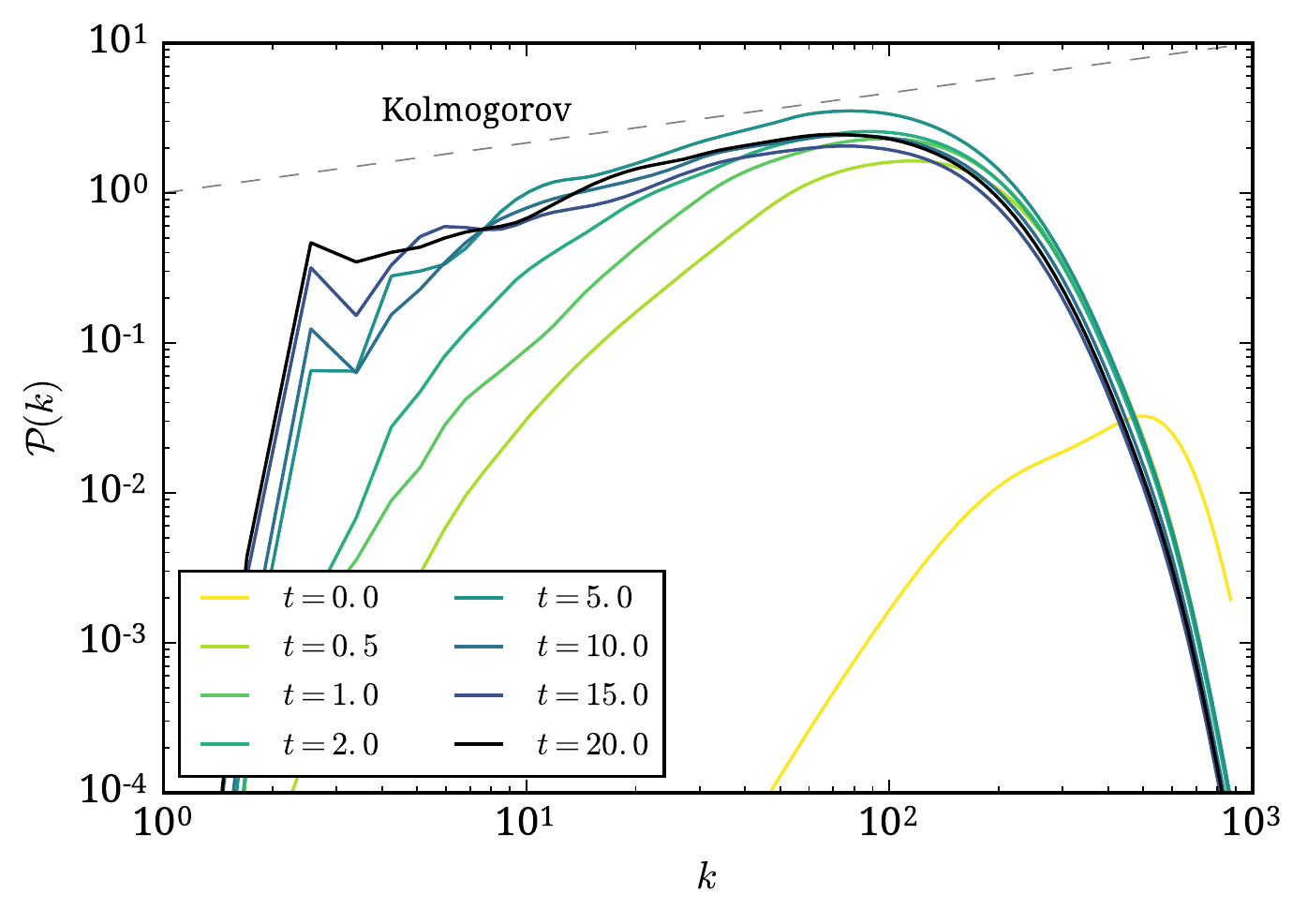}
\caption{Evolution of power spectra of velocity (left) and vorticity (right) for
model with the grid size $h = 1/1024$. We plot several power spectra calculated
at different times. \label{fig:spec_evol}}
\end{figure*}

In the bottom right plot of Figure~\ref{fig:vort_evol}, we show the mean inflow
speed $V_{in}$ measured at the upper and lower boundaries at $y = \pm2.0$. Since
these boundaries are open, the magnetic flux can be freely transported through
them. The positive values of $V_{in}$ indicate that the velocity at the boundary
is directed toward the interior of the domain, therefore, it brings fresh
magnetic flux to the system. In the standard Sweet-Parker model we interpret
this quantity as a measure of the reconnection process efficiency. The only way
magnetic field can be changed such system is by its conversion into other forms
of energy, dissipation, or transport through the open boundaries. Since our
inflow boundaries are far from the initial current sheet, we observe a zero
inflow during the initial period, followed by a very quick burst of the inflow
and then a decay. The maximum value of $V_{in}$ and its following decay seems to
be nearly independent of the grid size $h$. It indicates that the reconnection
process under the same physical conditions is independent of numerical
dissipation, even if its rate reaches values which are only small factors of
$V_A$. These small values of $V_{in}$ are mostly due to the fact, that our
boundaries along the X direction are periodic, so the reconnected flux is not
removed from the system. Comparing models with different sound speeds $a$ (or
plasma $\beta$ parameters) we see that the inflow speed $V_{in}$ is very
sensitive to $a$. The moment at which $V_{in}$ starts to quickly grow happens at
different times for models with different $a$, and $V_{in}$ reaches
significantly different values at later times. We verified that within the
turbulent region, the maximum sonic Mach number reaches around $2.0$ for model
with $=0.5$, and around $1.0$ or below $0.25$ for models with $a = 1.0$ and
$4.0$, respectively, signifying that the supersonic turbulence can be generated
in models with $a < 1.0$. This means that the compression may be ``squeezing''
more magnetic field within the turbulent region and/or it is more efficiently
dissipated in local shocks. This process would saturate in the presence of open
boundary conditions along the current sheet allowing for removal of the
reconnected magnetic field. However, with the periodic boundaries, the flux is
accumulated and therefore the inflow speed may not be a reliable measure of the
reconnection rate in these systems.

Figure~\ref{fig:vorticity_visualization} shows slices of total vorticity along
the three main mid-planes at three different moments, $t = 1.0$, $5.0$, and
$20.0$. All plots have the same color range for easier comparison and are
obtained from the model with the smallest grid size $h = 1/1024$. We see that
the stochastic reconnection develops complex filamentary structures near the
current sheet, increasing their volume with time, what supports the filling
factor evolution shown in the right plot of Figure~\ref{fig:vort_evol}. It is
important to recognize, that most of the turbulent motions are developed in the
ZY-plane perpendicular to the reconnecting field. Vorticity shows places in
which the velocity changes quickly. As we see in the middle column (the
ZY-plane), the high vorticity filaments have arbitrary orientations. In the
right plots (the XY-plane), however, the filaments tent to align with the X
direction, especially at later times (central and bottom plots). This is
explained well by the fact that the motions can mix reconnecting field lines
easier in the perpendicular plane than in the parallel direction, due to the
magnetic tension, even if these motions are produced by the reconnection of the
same lines. These visualizations demonstrate how a simple current sheet with a
weak velocity noise can create complex turbulent structure within a broad
vicinity of it.

%
\subsection{Turbulence Properties: Power Spectra and Anisotropy}

In the previous subsection we studied the evolution of turbulent motions
produced by stochastic reconnection and initiated from a random velocity noise.
We analyzed how the kinetic energy and vorticity evolves with time, what is the
growth rate of the turbulent region thickness and how the estimated reconnection
rate depends on the used grid size or $\beta$. Here we take a closer look into
the statistical properties of turbulent motions by studying their power spectra
and anisotropy. The fundamental question we want to answer here is if the
turbulence generated by reconnection can be characterized a by a mixture of
strong turbulence, described by \cite{GoldreichSridhar:1995} model, and
structures produced by the reconnection ejection regions.

\begin{figure*}
\centering
\includegraphics[width=0.32\textwidth]{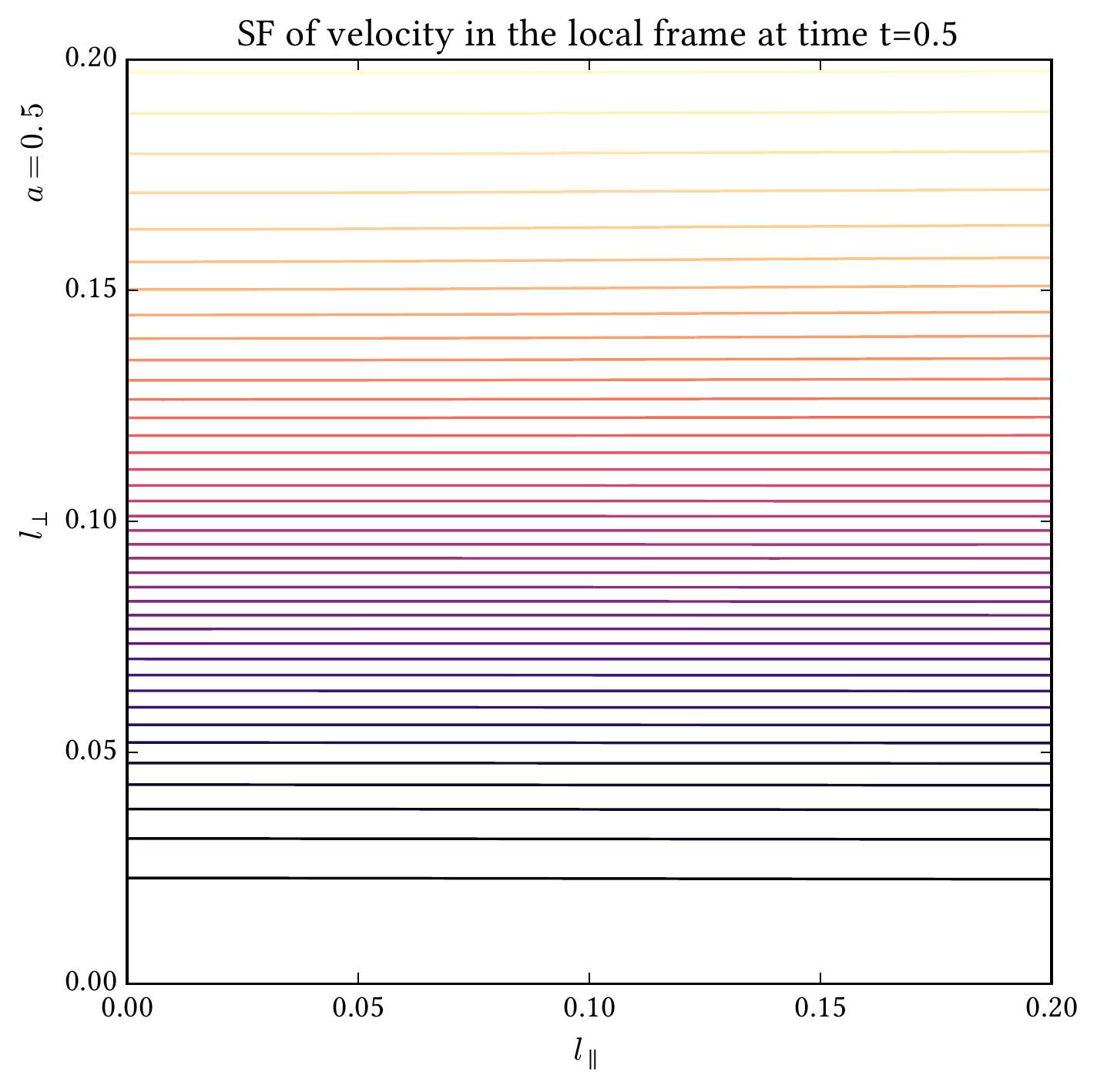}
\includegraphics[width=0.32\textwidth]{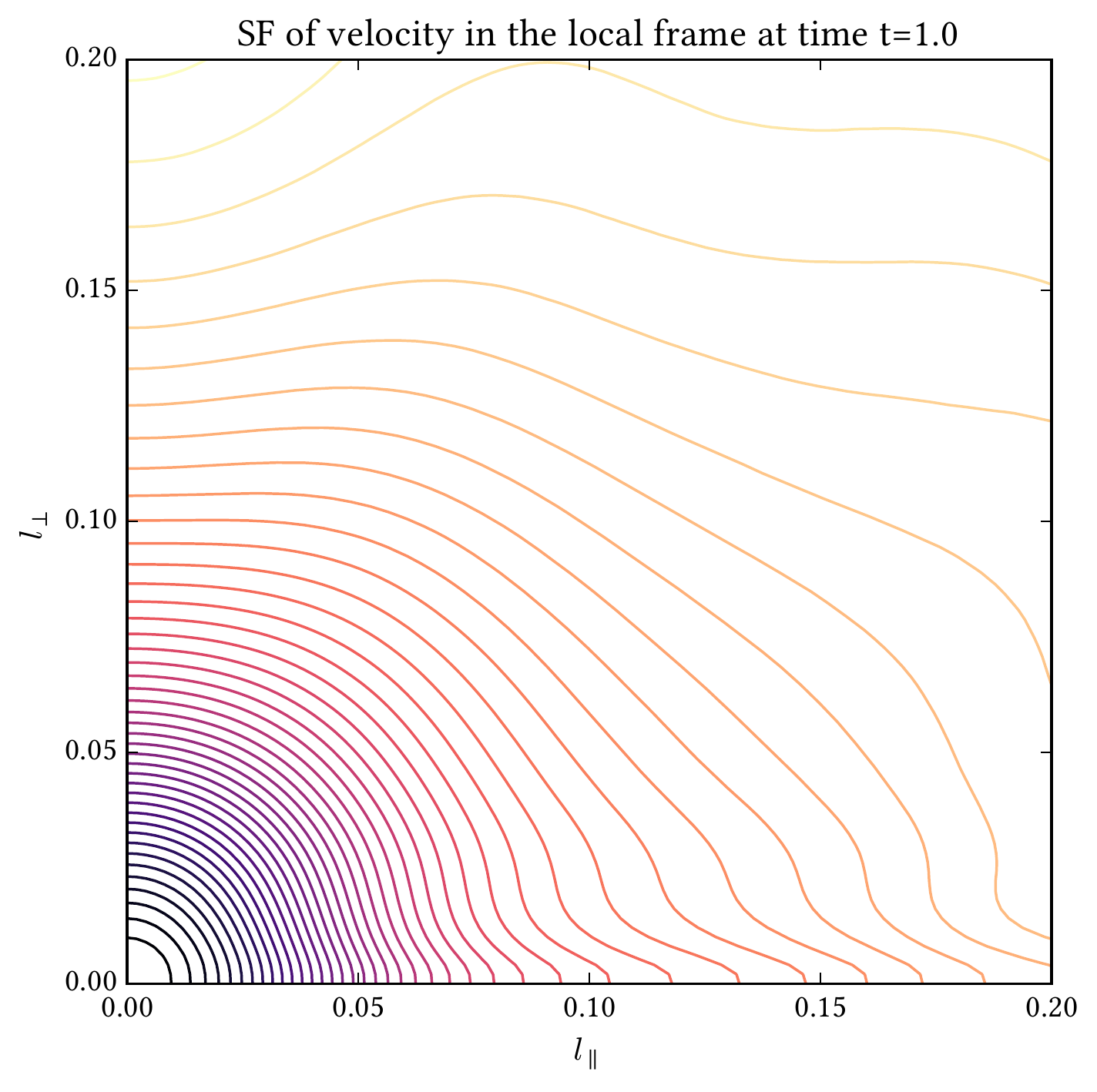}
\includegraphics[width=0.32\textwidth]{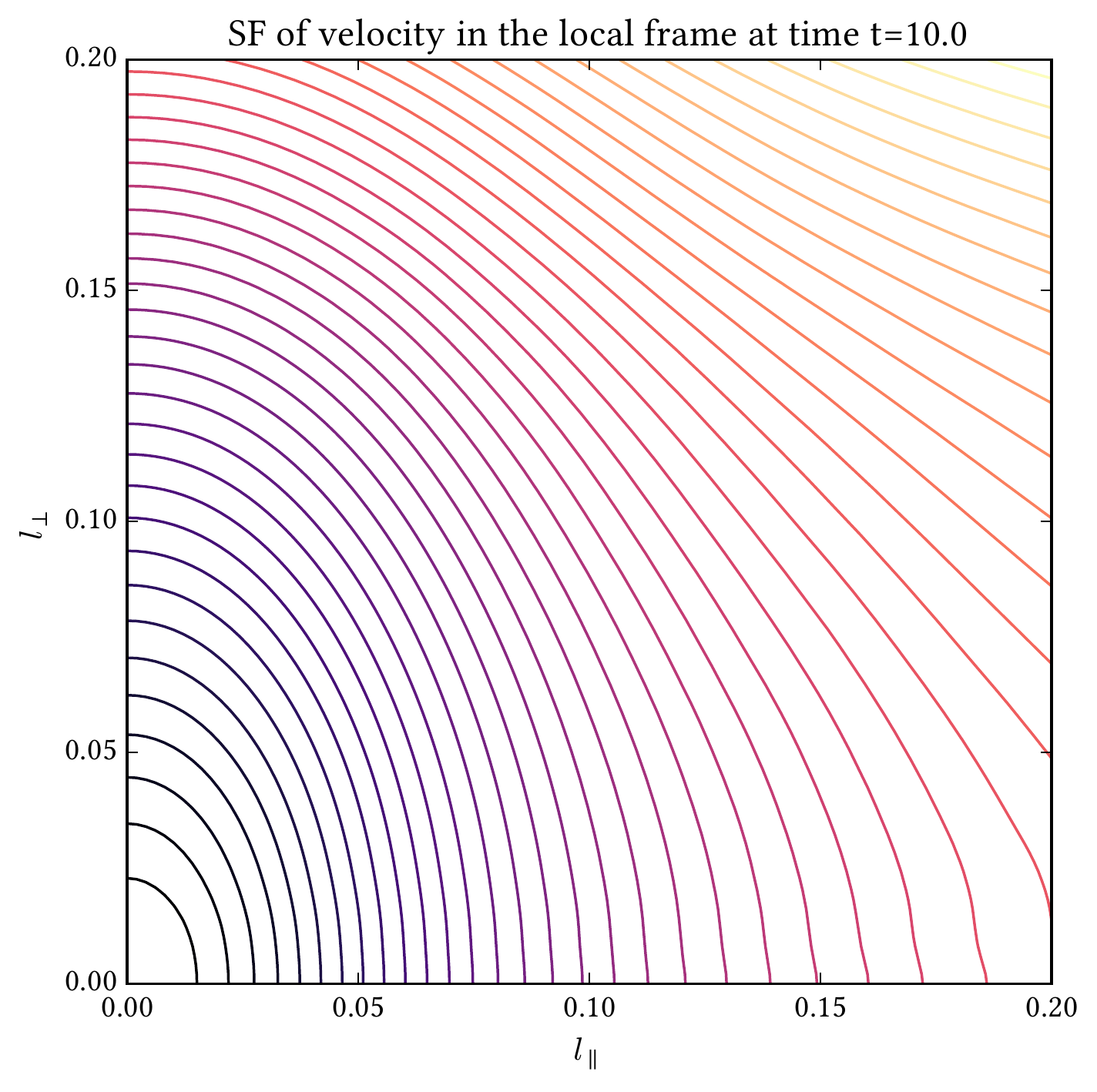} \\
\includegraphics[width=0.32\textwidth]{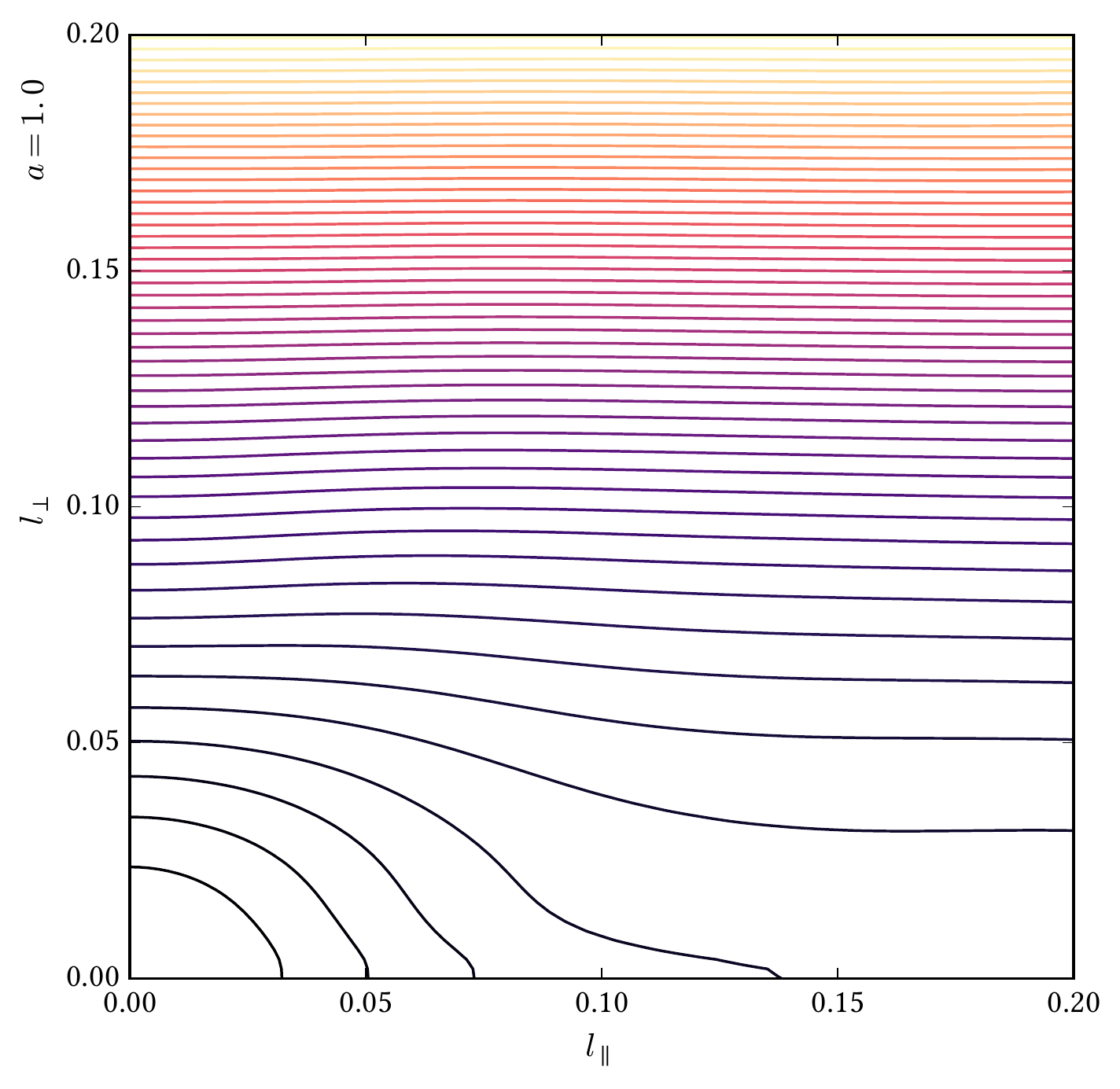}
\includegraphics[width=0.32\textwidth]{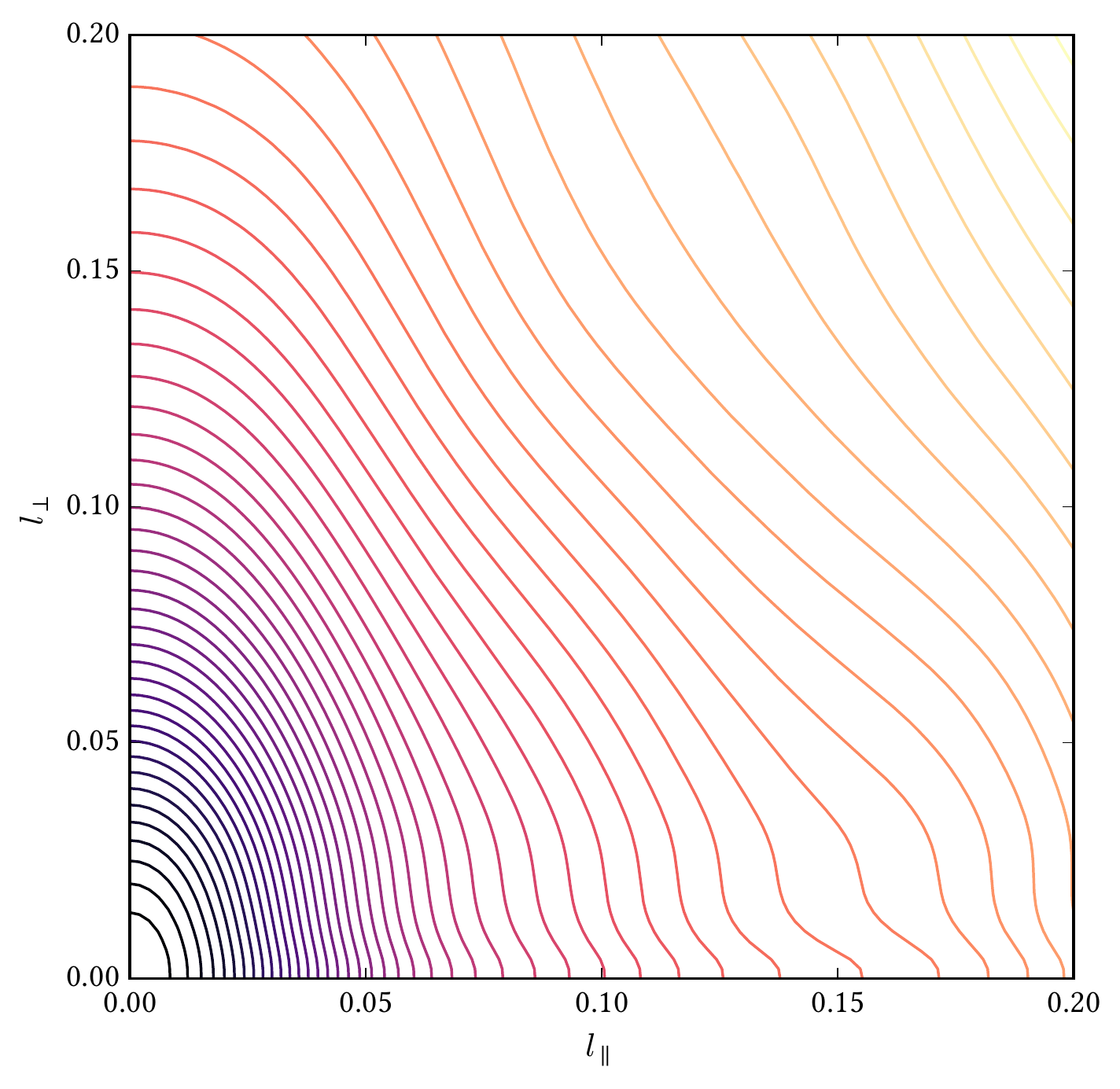}
\includegraphics[width=0.32\textwidth]{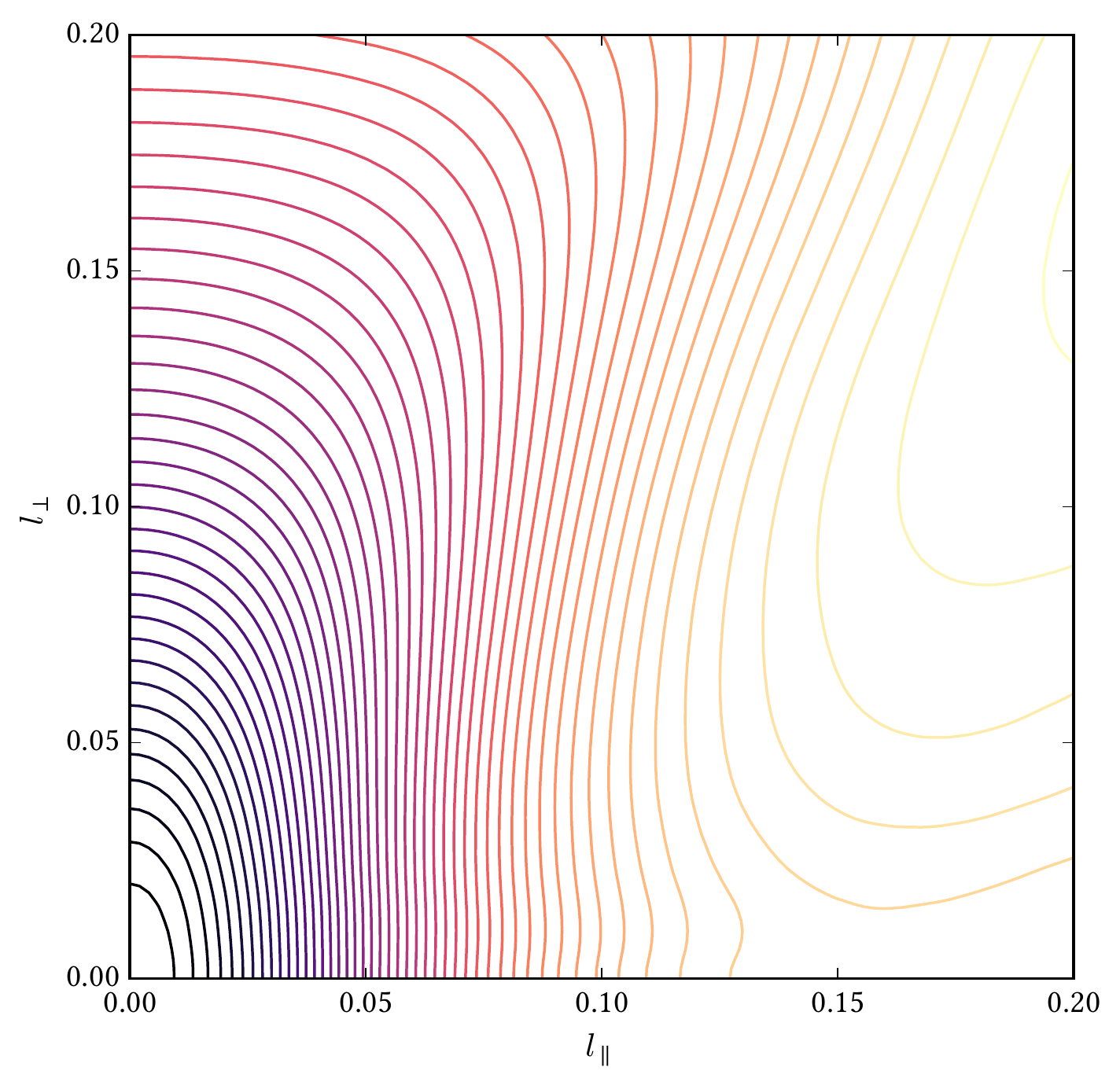} \\
\includegraphics[width=0.32\textwidth]{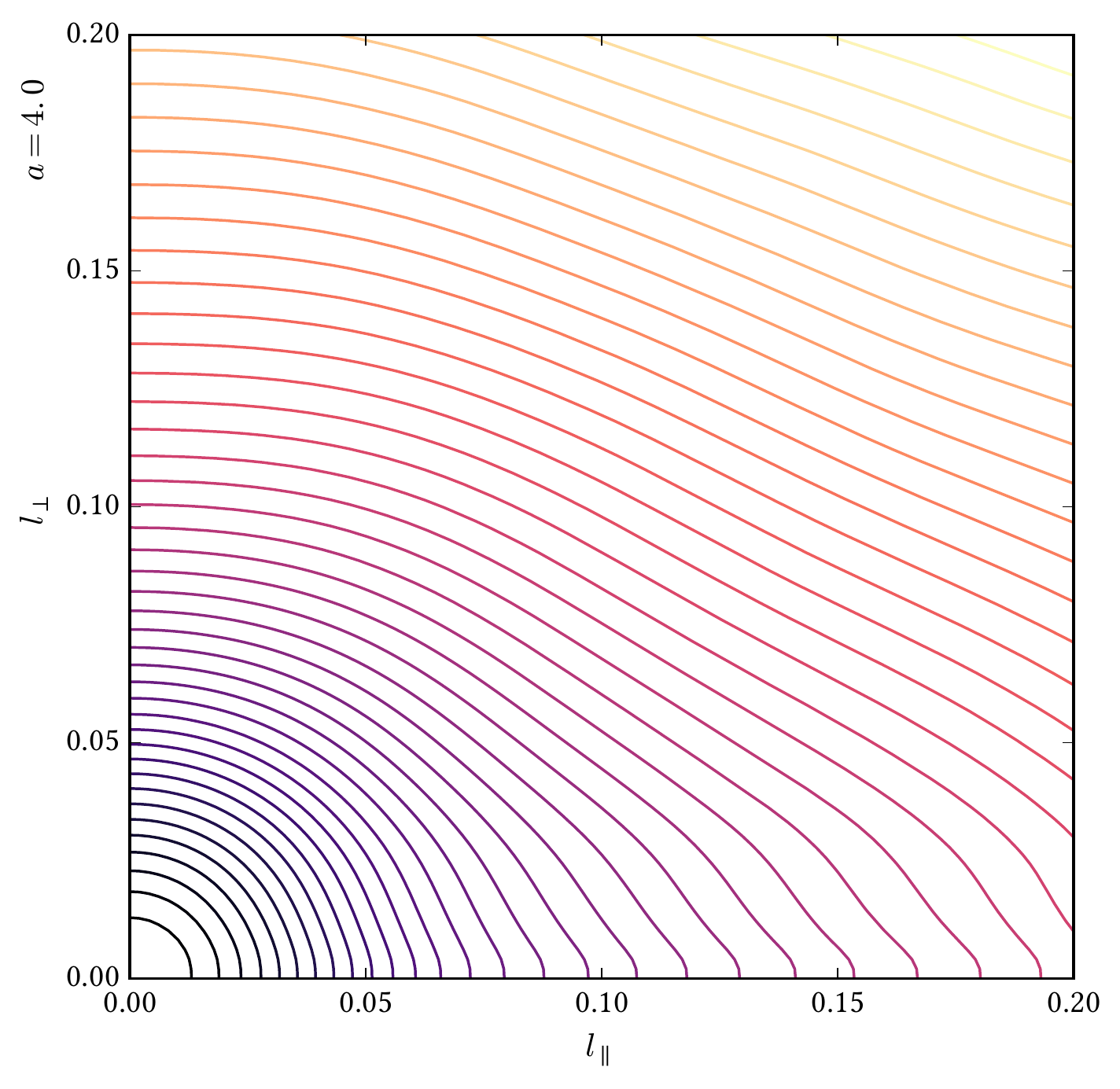}
\includegraphics[width=0.32\textwidth]{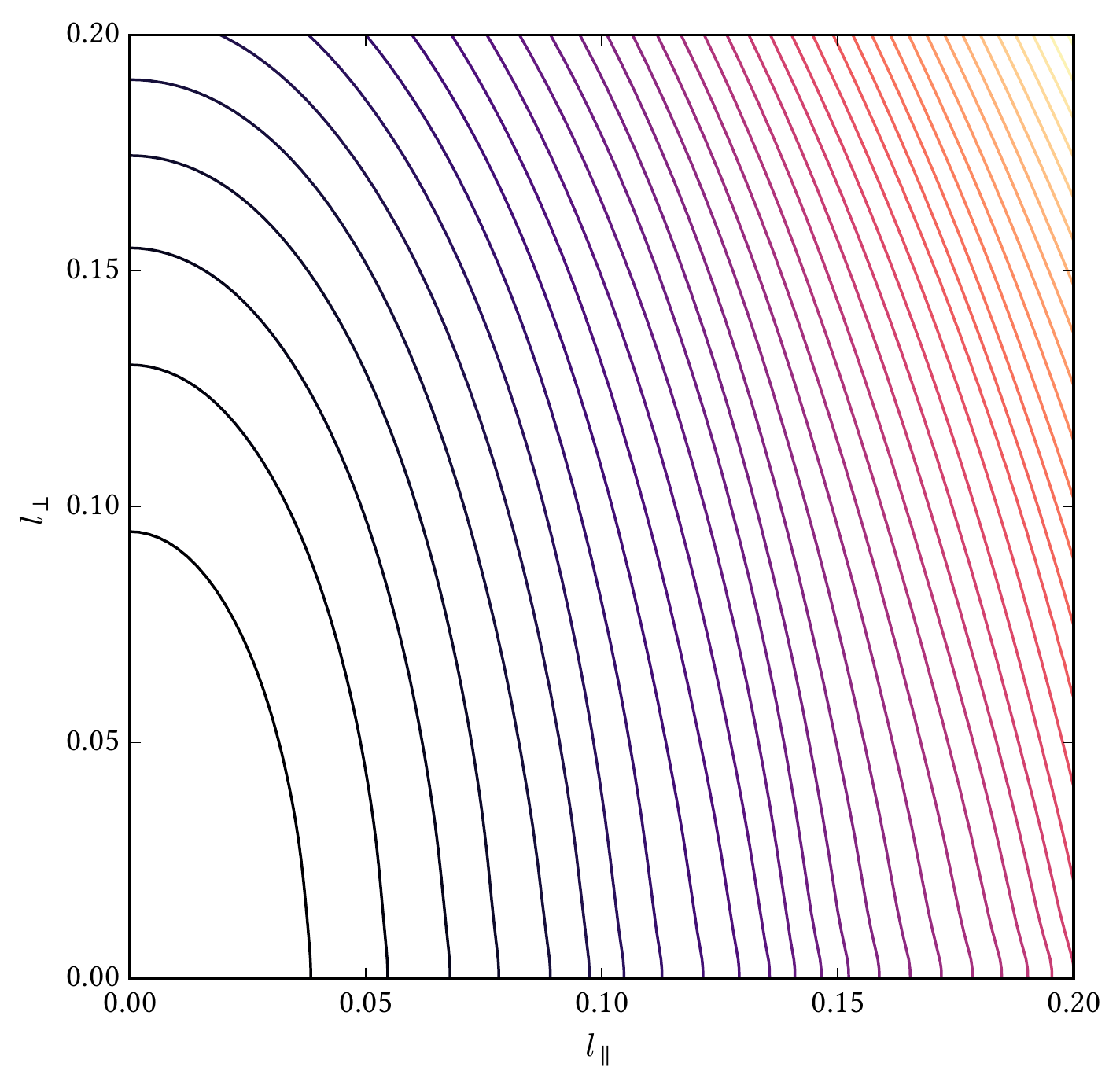}
\includegraphics[width=0.32\textwidth]{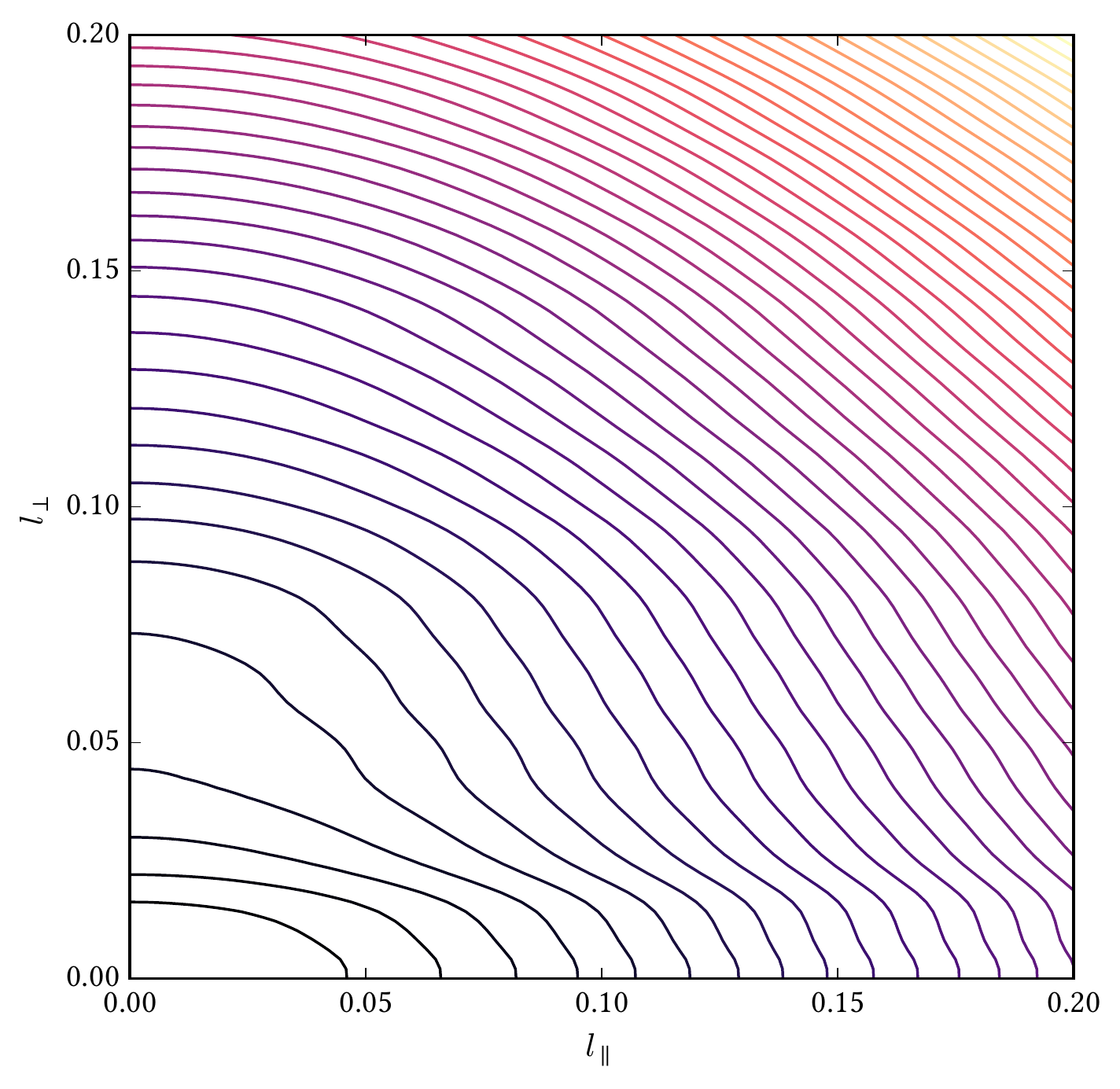}
\caption{Anisotropy of velocity structure functions with respect to the local
mean magnetic field at two early time moments $t=0.5$ (left) and $t=1.0$
(center), and later one at $t=10.0$ (right). Structure functions were calculated
for the model with the grid size $h = 1/512$ and box size $L_x = L_z = 1.0$. The
upper, middle and lower rows correspond to models with sound speed $a = 0.5$,
$1.0$ and $4.0$, respectively.
\label{fig:velo_anis}}
\end{figure*}

In Figure~\ref{fig:spec_evol} we show power spectra of velocity (left) and
vorticity (right) for different moments during the turbulence development. The
power spectra were calculated using 3D wavelets in Fourier space
\cite[see][]{Kirby:2005}. For comparison, we show Kolmogorov slope ($k^{-5/3}$
for velocity, and corresponding $k^{1/3}$ for vorticity) using gray dashed line.
The initial power spectrum ($t = 0.0$) is shown using the brightest line (yellow
in the color version). The velocity fluctuations are spread over the smallest
scales with very small amplitudes ($P(k) < 10^{-7}$ for velocity, see the left
plot). Fluctuations of both quantities, velocity and vorticity, quickly develop
from small to large scales. At $t = 0.5$, they are already spread over large
range of scales (up to $k \le 10$). After $t > 2.0$, power spectra start to
align with the Kolmogorov slope forming the inertial range. At $t = 5.0$ the
inertial range is well developed and it extends from $k \approx 10$ down to
nearly $k = 200$ for models with the effective grid size $h = 1/1024$. At the
final moment of simulation, $t = 20.0$, the power spectrum is fully developed
and stationary (compare its change for $t \ge 10$), both for velocity and
vorticity, and is characterized by a slope close to the Kolmogorov one.

The development of broad power spectrum from small to large scales cannot be
simply explained by reconnection which usually operates at small scales across
the current sheet. If considered one reconnection event, the ejection occurs
along a thin slab determined by the local current sheet thickness. However,
within the local ejection region, the magnetic field is oriented perpendicularly
to the current sheet plane, since the reconnected component is removed along the
slab. On the other hand, a typical fluctuation scale along the current sheet is
determined by the local separation of the reconnection events. Initially, this
separation scale is very short in our case due to densely packed reconnection
events, and later it increases with time due to the interactions between
ejections acting at larger and larger scales. This picture is well justified by
plots in the left column of Figure~\ref{fig:vorticity_visualization}, where a
typical scale of vorticity structures seem to increase with time (compare
ZX-plane cuts for different time moments). This also indicates a development of
velocity anisotropy dominated by initially small and nearly isotropic structures
and transformed into anisotropic ones. This picture can be confirmed if we
estimate the anisotropy with respect to the mean direction of reconnecting
components, i.e. along the X direction. However, what is the anisotropy with
respect to the local mean field and the reconnecting one? Should we expect two
different scalings, i.e. \citeauthor{GoldreichSridhar:1995}-like at large scales
and a different one produced by reconnection outflow regions at smaller scales?
Up to which scales the outflows can affect the statistics? How the anisotropy is
developed for different $\beta$-plasma parameters?

\begin{figure*}[t]
\centering
\includegraphics[width=0.48\textwidth]{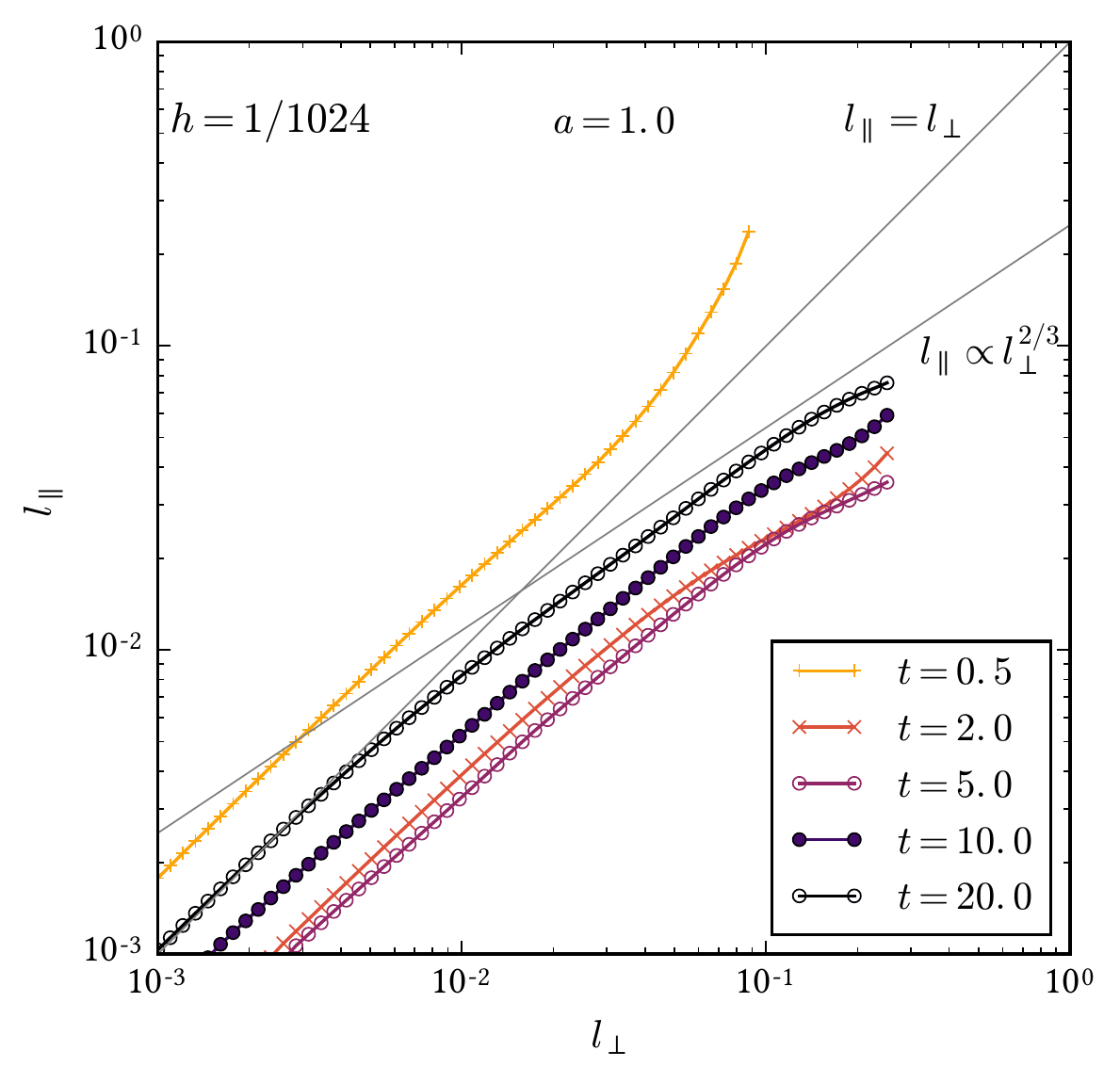}
\includegraphics[width=0.48\textwidth]{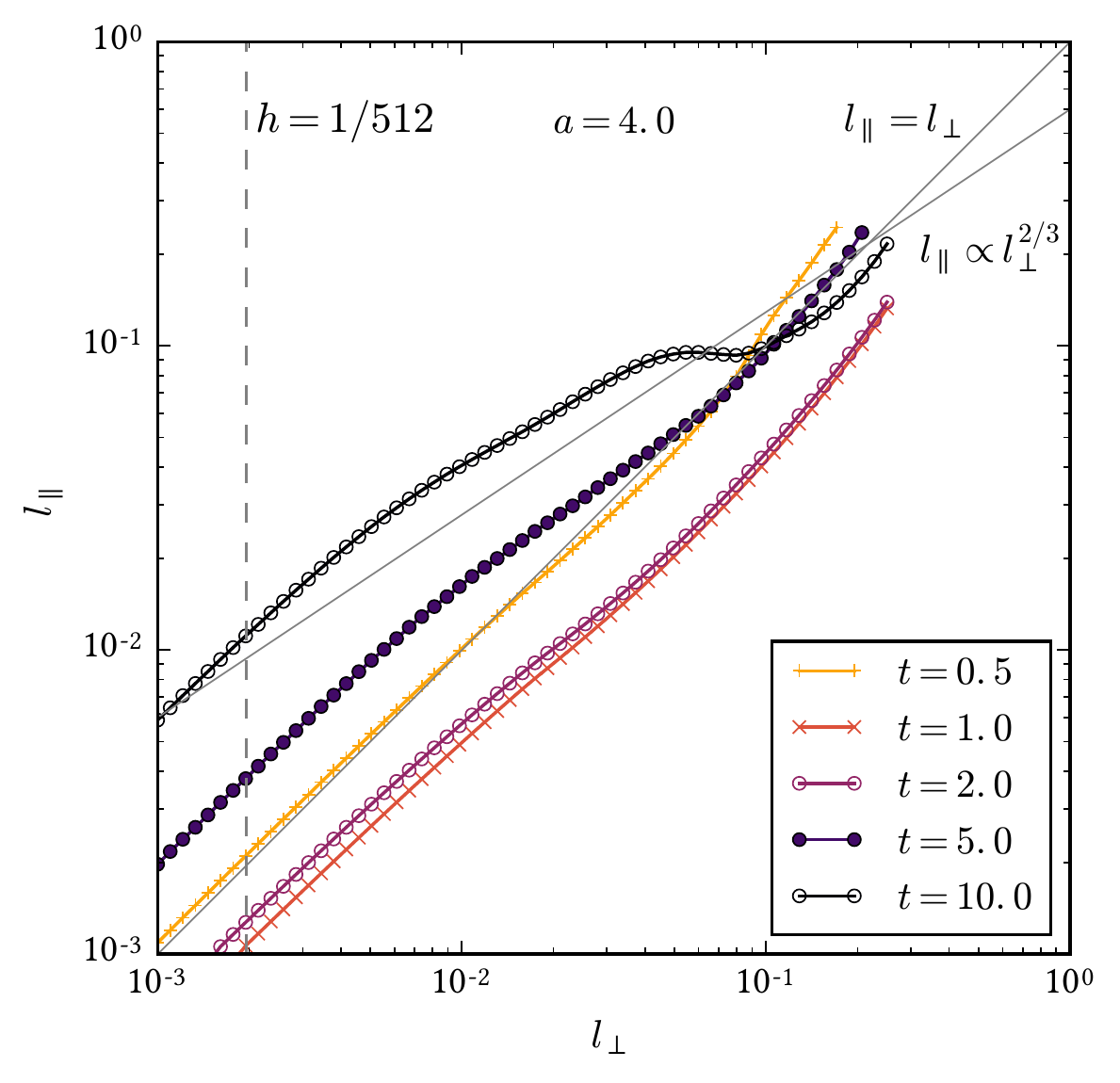}
\caption{Anisotropy scalings for total velocity for models with different sound
speeds ($a=1$ on the left and $a = 4$ on the right) obtained at different
moments. For comparison, we show isotropic and strong turbulence scalings
(denoted by gray thin lines). The vertical dashed line in the right plot shows
the effective grid size $h = 1/512$. In the left one, the grid size is below the
left axis.
\label{fig:velocity_anisotropy}}
\end{figure*}

In Figure~\ref{fig:velo_anis} we show the second-order structure functions of
velocity calculated in the local reference frame, i.e. with respect to the local
mean magnetic field at three different time moments: $t = 0.5$, $1.0$, and
$10.0$ (left, center, and right columns, respectively) for models with $a =
0.5$, $1.0$, and $4.0$ (upper, middle and lower rows, respectively). We notice
the change of anisotropy degree $l_\parallel / l_\perp$ between different
moments. For all three models with different $a$ at time $t = 0.5$ the velocity
fluctuations are elongated with the local magnetic field. We clearly see that
the anisotropy decreases with increasing $a$ (or $\beta$). However, once the
reconnection increases its efficiency, the velocity fluctuations start to be
elongated rather in the direction perpendicular to the local field (see the
central column). At $t = 1.0$, the velocity fluctuations are relatively
isotropic for model with $a = 0.5$ (for $l < 0.1$), while for other models they
are elongated in the perpendicular direction to the local field.

Why this sudden change of orientation of anisotropic velocity structures? Our
understanding is that the Alfv\'en waves dominate initially the structure of
velocity, but the local reconnection events become quickly stronger taking over
the statistics. A local reconnection event can be characterized by a weak and
broad inflow of the magnetic flux, and strong and thin outflow region (see the
typical Sweet-Parker picture). Within the outflow region, the velocity is
oriented in the direction perpendicular to the local field (i.e. with respect to
the reconnected flux component and the guide field), therefore the flow is
perpendicular to this field. This is clearly observed in the right top plot of
Figure~\ref{fig:velo_anis} showing the local structure function at $t = 10.0$,
where its value increases quickly along the parallel direction, determining the
averaged thickness of the current sheet ($\delta < 0.05$), while in the
perpendicular direction it increases slowly up to a fraction of the length unit.
This anisotropy continues to dominate in perpendicular direction during the
later times of our simulation (see the right plots of Figure~\ref{fig:velo_anis}
for $a = 0.5$ and $1.0$). At later time $t = 10.0$ for the model with the
highest sound speed ($a = 4.0$) the anisotropy degree indicates that
fluctuations change their orientation and are aligned with the local magnetic
field rather, on the contrary to models $a \le 1.0$. This indicates that the
compressibility could be an important factor in determining the small scale
anisotropy in reconnection-driven turbulence, since it increases with decreasing
sound speed.

As we discussed above, the reconnection generate strong anisotropy in the
perpendicular direction to the local field, and during the evolution the
anisotropy degree changes significantly. The dominance of the perpendicularly
oriented fluctuations is observed in the whole evolution of models with $a \le
1.0$ and during the initial few Alfv\'en time units in the model with $a = 4.0$.
The difference comes from the fact, that the reconnection outflow interactions
decay to turbulent fluctuations, and the timescale for this process seem to
depend on fast magnetosonic speed, since it is faster for higher sound speeds.
Of course, this results also in the same timescale for anisotropy changes. The
perpendicular anisotropy dominance is in contrast to the Goldreich-Sridhar
turbulence where velocity eddies are elongated with the magnetic field. We
explained, that the difference comes from the strong reconnection outflows
dominating the structure function, especially at small scales. However, even
though the fluctuations are not oriented along the local field, maybe at least
the scaling of anisotropy reveals some compatibility with the Goldreich-Sridhar
model?

In Figure~\ref{fig:velocity_anisotropy} we show anisotropy scaling of velocity
fluctuations at several moments, $t = 0.5$, $2.0$, $5.0$, $10.0$, and $20.0$ for
the model with $a = 1.0$ ($h = 1/1024$, left plot), and $t = 0.5$, $1.0$, $2.0$,
$5.0$, and $10.0$ for the model with $a = 4.0$ ($h = 1/512$, right plot). These
plots confirm that the anisotropy degree varies significantly during the system
evolution, especially at the beginning of the simulation. Initially isotropic
fluctuations develop strong anisotropy at all scales along the perpendicular
direction $l_\perp$ (shown in the horizontal axis) with scaling close to
$l_\parallel \propto l_\perp$, but after about $t = 5.0$ for the model with $a =
1.0$ and $t = 2.0$ for the model with $a = 4.0$ the anisotropy degree tends to
decrease. Even though the anisotropy changes significantly, its scaling seems to
be relatively consistent up to $t = 10.0$ for the model with $a = 1.0$. We
observe that for scales below $0.1$ it continues to follow $l_\parallel \propto
l_\perp$, while only at larger scales it starts to manifest the
Goldreich-Sridhar scaling $l_\parallel \propto l_\perp^{2/3}$. On the right, the
anisotropy scaling becomes compatible with $l_\parallel \propto l_\perp^{2/3}$
earlier. We see it is formed already at $t = 5.0$ from $l_\perp \approx 0.1$
down to scales as small as $0.01$. At this time, the turbulent region is
relatively thick, as estimated in the bottom left plot in
Figure~\ref{fig:vort_evol}. The developed turbulence have enough space to create
inertial range and structure compatible with strong turbulence. At the same time
reconnection events spread sparsely over the same turbulent region, affecting
only the local velocity structures and relaxing to turbulent motions at shorter
scales for the model with higher sound speed. This indicates the presence of
strong turbulence described by the Goldreich-Sridhar theory, substantially
disturbed by the injection from the reconnection outflows at smaller scales,
characterized by different properties.

%
\subsection{Decaying Turbulence without Global Field Reversal}

In order to remove the strong contribution of the reconnection outflows observed
in the anisotropy scaling of velocity fluctuations, which is responsible for
deviation from \cite{GoldreichSridhar:1995} scaling in the model with sound
speed $a = 1.0$, we performed a run with the grid size $h = 1/512$ in which we
removed the global reversal of magnetic field by setting $|B_x|$ everywhere and
leaving other two components, $B_y$ and $B_z$, unchanged. From such initial
state, we restarted our simulation and let the fluctuations evolve without
energy injection from the reconnection process. Due to the lack of global
reversed field, the reconnection should be strongly suppressed. We should note
that the reconnection can still produce weak energy injection from the unchanged
components of magnetic field. Nevertheless, the main source of injection should
be removed, and previously produced velocity and magnetic field fluctuations
should start to decay due to the dissipation. We expect, that due to the lack of
reconnection outflows, the fluctuations should quickly produce scaling
compatible with the Goldreich-Sridhar model. We ran this model for a few
Alfv\'en times, and then calculated local structure functions in order to
analyze the velocity anisotropy scaling.

\begin{figure}[t]
\centering
\includegraphics[width=0.48\textwidth]{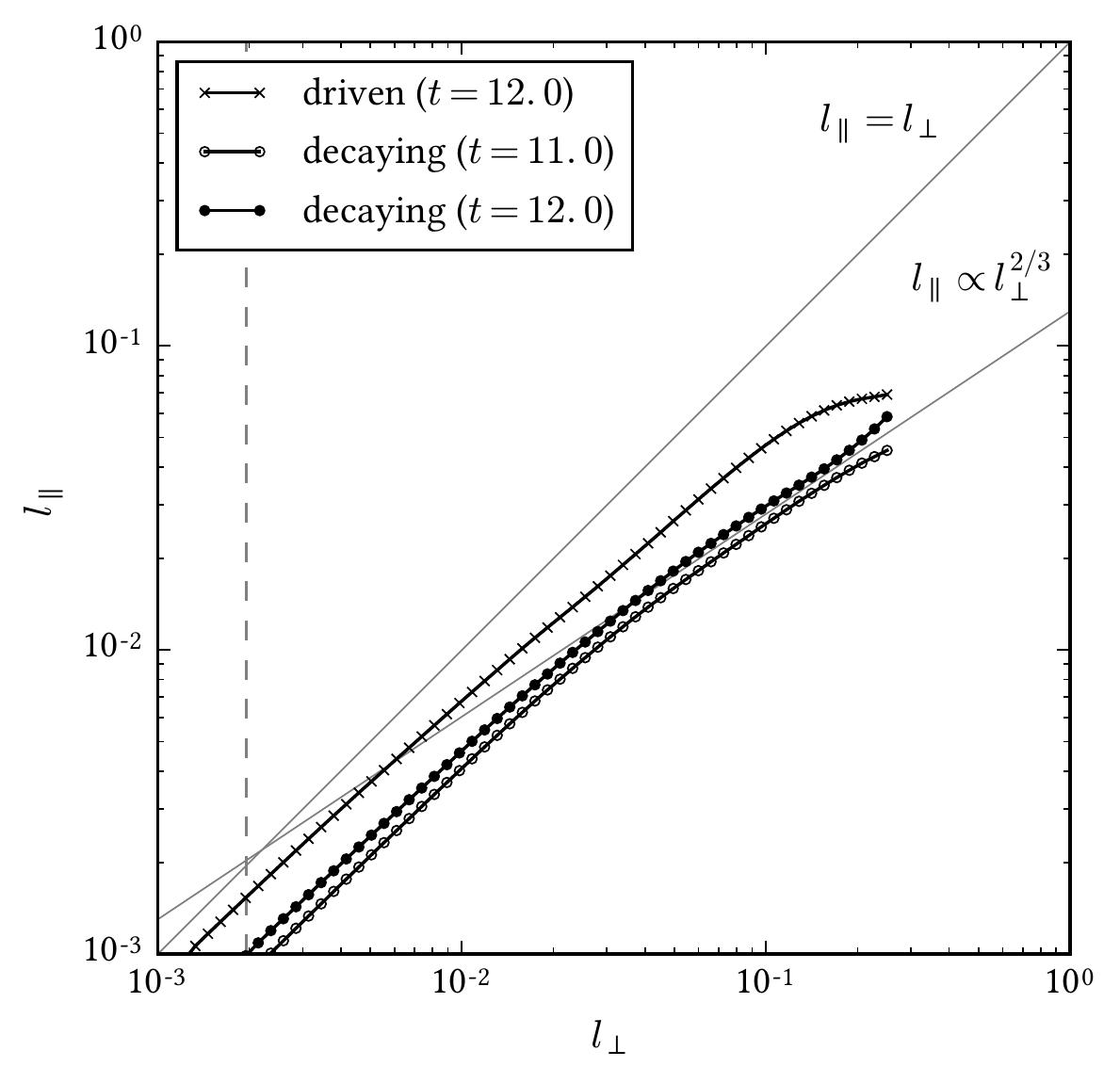}
\caption{Comparison of the anisotropy scalings for velocity in the
reconnection-driven turbulence run with the global field reversal at $t = 12.0$
($\times$ points) and at two time moments $t = 11.0$ and $12.0$ (open and closed
circles, respectively) for model in which the global reversal of magnetic field
was removed at $t = 10.0$ and the simulation was restarted. We see that while in
the reconnection-driven run the reconnection outflows affect the velocity
statistics and the scaling is closer to the isotropic one, in the run without
strong outflows (no field reversals) the anisotropy scaling tends to the
\cite{GoldreichSridhar:1995} scaling, $l_\parallel \propto l_\perp^{2/3}$, at
scales $l_\perp > 2\cdot10^{-2}$.
\label{fig:decaying_velocity_anisotropy}}
\end{figure}

In Figure~\ref{fig:decaying_velocity_anisotropy} we compare anisotropy scalings
for driven and decaying models at the same moment $t = 12.0$, i.e. the
reconnection-driven turbulence shown with $\times$-points, and the decaying
turbulence in which the global magnetic field reversal was removed (open and
closed circles for $t = 11.0$ and $t = 12.0$, respectively). We clearly see the
difference in the anisotropy scaling. In the reconnection-driven case, the
outflows still strongly affect the velocity statistics and a scaling closer to
$l_\parallel \propto l_\perp$ is observed. In the case of decaying turbulence,
the reconnection outflows do not affect the statistics anymore and after one
Alfv\'en time the scaling is already aligned with $l_\parallel \propto
l_\perp^{2/3}$ at larger scales. The same is observed at later time $t = 12.0$.
This test indicates, that the reconnection outflows are responsible for the
change of anisotropy scaling.

%
\section{Discussion}
\label{sec:discussion}

\subsection{Comparison to previous results}

We performed numerical modeling of reconnection-driven turbulence and studied
its properties. We should note that we did not impose an initial Sweet-Parker or
Harris current configurations \cite[as in][]{Lapenta:2008, Oishi_etal:2015,
HuangBhattacharjee:2016}, but we let a weak noise affect the current sheet along
which magnetic field is discontinuous initially. Moreover, in this work we
applied periodic boundary conditions along the current sheet, but we left the
vertical boundaries open. With the exception of the vertical boundary type, our
setup resembles closely the one studied by \cite{Beresnyak:2013}.

The choice of different vertical boundaries have some important consequences.
Periodicity in all three directions with two separated current sheets imposed,
as used in \cite{Beresnyak:2013}, allows for the possibility of horizontal large
scale interactions through the deformations of both current sheets. These
interactions, especially at later times, can generate large scale motions in the
perpendicular direction to the current sheet providing additional to the
reconnection energy input\cite[see][for a similar setup with many current
sheets, in which the current sheet deformations are well
manifested]{Drake_etal:2010, Kowal_etal:2011}. In our case, such large-scale
interactions are not allowed, therefore the only energy input comes from the
small-scale reconnection events. The reconnection rate values obtained from our
models are compatible to those in \cite{Beresnyak:2013}. We estimated the
reconnection rate to be about $0.025-0.035 V_A$, while the value estimated in
\cite{Beresnyak:2013} is around twice smaller. This can be justified by less
diffusive spectral code together with smaller resistivity coefficient (of the
order of $4\cdot10^{-5}$, as estimated from the Lundquist number for their
lowest resolution run) used in their work. The spectral slope of the turbulence
generated in their models was estimated to be around Kolmogorov slope, which is
comparable with the one we show in Figure~\ref{fig:spec_evol}.
\cite{Beresnyak:2013} did not demonstrate any results related to the properties
of velocity fluctuations, however, and since they focused only on incompressible
MHD regime, they did not study $\beta$-dependence.

\begin{figure}[t]
\centering
\includegraphics[width=0.48\textwidth]{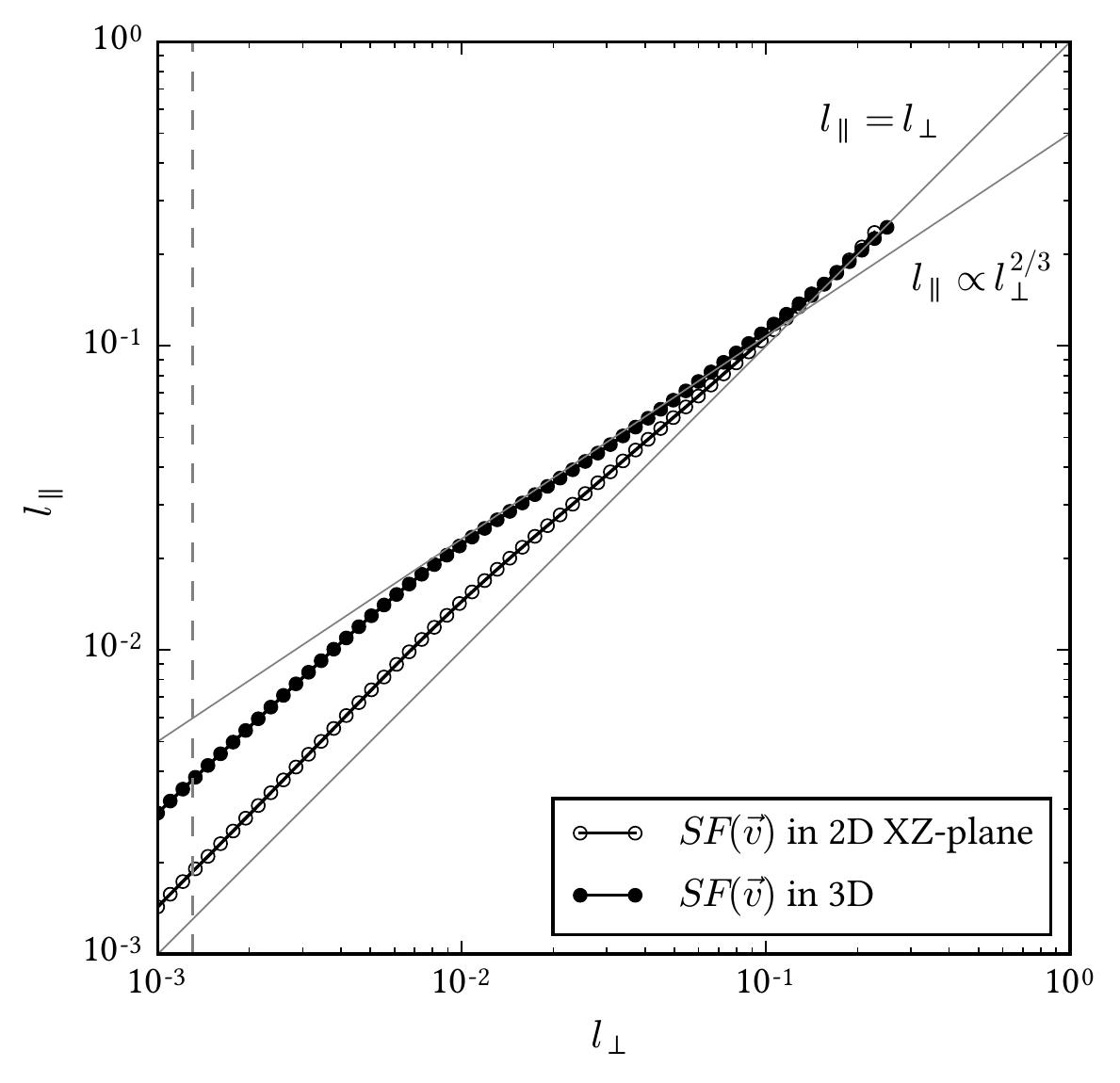}
\caption{Comparison of the anisotropy scalings for velocity in 3D driven
turbulence model without the global field reversal obtained from 3D structure
function calculated in the local reference frame (solid circles), and the
structure functions calculated in 2D XZ-planes with the local reference frame
defined by the in-plane field (open circles). In the 3D structure function we
see perfect scaling compatible with \cite{GoldreichSridhar:1995} model from
scales $l_\perp$ below $10^{-2}$ to above $10^{-1}$. Using the structure
function calculated in the XZ-plane with the in-plane defined local reference
frame, this scaling is lost. The vertical dashed line corresponds to the grid
size $h = 1/768$. A fully developed turbulent realization at $t = 13.0 t_A$ was
used.
\label{fig:turbulent_velocity_anisotropy}}
\end{figure}

As we mentioned already, \cite{Oishi_etal:2015} used a different numerical setup
starting from the Sweet-Parker configuration and periodic box along the current
sheet. The Sweet-Parker configuration requires open boundary conditions to allow
inflow and outflow of the magnetic flux in order to maintain steady state. Since
the periodic boundary conditions were applied, the growth of the fluctuations at
large scales may be attributed to the presence of the configuration of magnetic
field which is not in initial equilibrium. It is also important to notice that
these authors have misleadingly claimed to perform very long simulations up to
$1000 t_A$. They assumed an uncommon definition of Alfv\'en time $t_A = \delta_0
/ V_A$, where $\delta_0 = 0.02$ was the current sheet thickness. Typically, the
timescale is defined by $t_A = L / V_A$, where $L$ is the size of the system,
which results in their maximum simulation time of $20 t_A$, same as in our
models. Another important conclusion claimed by \cite{Oishi_etal:2015} was that
for sufficiently large Lundquist number $S$, the $\delta_{SP}$ becomes weakly
dependent on $S$ (see their Figure~1). However, the authors did not provide any
estimation of the numerical resistivity $\eta_{num}$, which we expect to be
comparable with the one we estimated, since both studies were performed using
Godunov-type codes \footnote{In \cite{Kowal_etal:2009} we estimated that
$\eta_{num} \approx 10^{-4}$ for models with the grid size $h=1/512$.}. From
Table~1 in \cite{Oishi_etal:2015} we can see that for high Lundquist number
models, the explicit resistivity $\eta$ is much below $\eta_{num}$ we estimated.
Therefore, the deviation from Sweet-Parker dependence $V_{rec} \propto S^{-1/2}$
observed in their simulation could be artificial and resulting from the
numerical effects.

Recently, \cite{HuangBhattacharjee:2016} analyzed the statistics of velocity
fluctuations generated by reconnection. They also used as a starting point a
global Sweet-Parker configuration, however, with non-periodic boundaries along
the current sheet. Weak velocity perturbations were injected into such domain in
order to understand their effect on the reconnection rate. If the perturbations
are injected too far from the current sheet, they simply propagate out of the
domain as Alfv\'en waves before reaching the reconnection zone. They used a
guide field comparable in strength to the reconnecting component, which
additionally prevents the interactions of the waves. They calculated windowed 2D
power spectra of velocity over $XZ$-planes and averaged them over the vertical
direction. Their power spectra demonstrated steeper slope comparing to
Kolmogorov turbulence. The structure functions were also calculated over
$XZ$-planes, and the local reference frame was determined by the in-plane
magnetic field. Here, we applied the same technique for calculation of local
structure function in turbulent simulations, and found that this method results
in much reduced anisotropy, when compared to the fully 3D structure function
analysis. In Figure~\ref{fig:turbulent_velocity_anisotropy} we present the
comparison between both methods applied to driven turbulence model without
global field reversal simulated in fully periodic box with the grid size $h =
1/768$. The reduction of the anisotropy scaling due to a different technique is
clear. This indicates, that these authors' conclusion of a different type of
turbulence (steeper power spectrum and different than Goldreich-Sridhar
anisotropy scaling) should be taken carefully. In our numerical experiments we
obtain Kolmogorov-like power spectra of generated velocity fluctuations, which
also present Goldreich-Sridhar anisotropy scaling at later times in large
$\beta$ models. This indicates that the turbulent statistics are similar to
strong MHD turbulence, but are strongly affected (contaminated) by the dynamics
of reconnection induced flows. In particular this is clearly visible for low
$\beta$ models, where supersonic reconnection outflows are present. In high
$\beta$ models we observe that the Goldreich-Sridhar anisotropy scaling is
visible at much earlier times, since the fluctuations generated by reconnection
outflows are propagated faster. We also claim that the Goldreich-Sridhar
anisotropy scaling is more apparent for models with higher resolution due to
larger dispersion in scales between the turbulent fluctuations and reconnection
outflow, what is confirmed by Figure~\ref{fig:velocity_anisotropy}. As a
consequence, our conclusions differ from those of \cite{HuangBhattacharjee:2016}
with regards to the type of turbulence in magnetic reconnecting plasmas.

\subsection{Validity of our approach}

In this paper we study magnetic reconnection using magnetohydrodynamic (MHD)
description of a plasma (see Section~\ref{sec:model}). The validity of this
description has been broadly discussed in literature \cite[see
e.g.][]{Kulsrud:1983, Somov:2006a, Eyink_etal:2011, Lazarian_etal:2015} and its
justification can be based on either collisionality or strong magnetization.
Following \cite{Eyink_etal:2011}, where we refer for more detailed review of the
validity of MHD description, we consider three characteristic length scales of
importance: the ion gyroradius $\rho_i$, the ion mean-free-path length
$\lambda_{{\rm mfp},i}$ due to Coulomb collisions, and the scale $L$ of
large-scale variations of magnetic field and velocity. The fluid picture of a
plasma is justified if the plasma is at least somewhat collisional, i.e. $L \gg
\lambda_{{\rm mfp},i}$. This regime can be divided into ``weakly collisional''
($\lambda_{{\rm mfp},i} \gg \rho_i$) and ``strongly collisional'' ($\rho_i \gg
\lambda_{{\rm mfp},i}$) plasmas. In the opposite case, i.e. $\lambda_{{\rm
mfp},i} \gg L$, the plasma is ``collisionless'', and hydrodynamic description
does not work anymore. In astrophysical plasmas we can encounter all three
regimes of collisionality. For example, plasmas in star interiors and accretion
disks are strongly collisional. Very hot and diffuse plasmas, such as warm
ionized interstellar medium, are weakly collisional, while solar wind at
magnetosphere or post-coronal mass ejection (CME) current sheets are examples of
collisionless plasmas \cite[see Table~1 in][]{Eyink_etal:2011}. Among the
additional assumptions applied to MHD description is the assumption of high
conductivity, which is easily fulfilled for most astrophysical plasmas \cite[see
e.g.][]{Somov:2006b}.

In this work we are interested in reconnection processes at length scales much
larger than $\rho_i$. For such scales plasmas can be more precisely described by
``kinetic MHD equations'', which differ from the standard MHD equations by the
isotropic thermal pressure $p$ replaced with the pressure tensor, which has two
components, $p_\parallel$ and $p_\perp$, parallel and perpendicular to the local
magnetic field, respectively \cite[see e.g.][]{Chew_etal:1956}. In the presence
of anisotropic pressure, plasma can develop kinetic instabilities, such as
``firehose'' or ``mirror'' instabilities, which strongly affect the plasma
evolution at small scales \cite[see e.g.][ and references
therein]{HauWang:2007}. Our focus in the studies here is to prohibit development
of any microscopic instabilities (by using the Particle-in-Cell or kinetic MHD
approaches) or enhanced reconnection rates (by using Hall term or anomalous
resistivity), which could influence the generation of turbulence by reconnection
process, even if other frameworks describe better the astrophysical plasma of
interest in which the reconnection process takes place.

\subsection{Restrictions imposed by periodic boundary conditions}

As we mentioned in Section~\ref{sec:model}, the choice of periodic boundary
conditions along the X and Z directions imposes some restrictions on the
applicability of the model. For instance, the fluctuations of developed
turbulence in the parallel direction to the initial current sheet (the XZ-plane)
cannot reach sizes larger than the computational domain, therefore after around
$t \approx t_A = V_A^{-1}$, the energy could be accumulated at large scales due
to the interactions with the domain boundaries and enforce the inertial range
production down to the small scales. This, as result, may affect the development
of anisotropy. On the other hand, dissipation of the cascade originated waves
occurs in similar timescales. One would expect then the energy at small scales
to be greatly dominated by the local cascade and less affected by the boundary
effects. We performed models with different ratios of the Z to X dimensions of
the box ($L_z = 0.5 L_x$ versus $L_z = L_x$) and do not see significant changes
in the developed spectra and anisotropy statistics. The problem of periodic
boundary conditions, however, must be properly addressed and quantified in a
follow-up paper.

\subsection{What drives the observed turbulence?}

One of the most important results of this work is the reassurance of
self-generation of turbulence in reconnection events. The other is that this
turbulence follows standard Kolmogorov and Goldreich-Sridhar statistics.
However, a still open issue is related to the driving mechanism of the observed
turbulent motions. What drives the turbulence in reconnection events? Literature
has suggested, without any quantitative proof, that tearing modes, plasmoid
instabilities, and shear-induced instabilities could mediate the energy transfer
from coherent to turbulent flows.

In our numerical experiments we do not identify tearing modes, although
filamentary plasmoid-like structures are present. The filling factor of these
are, however, visually recognized as very small. Sheared flows, on the other
hand, are present around and within the whole current sheet. As the field lines
reconnect the ${\bf v} \times {\bf B} + {\bf E}$ force increases, accelerating
the plasma and creating the current sheet. This process is, in three dimensions,
patchy and bursty. Therefore, the accelerated flows are strongly sheared. The
statistical importance of these burst flows is large, as already shown in this
work, as we compared the velocity anisotropy of reconnecting events to that of
decaying turbulence without the reversed field. Kelvin-Helmholtz instability due
to the sheared velocities in reconnecting layers has already been conjectured as
possible origin of turbulence by \cite{Beresnyak:2013}. In this work, we
therefore provide real evidence for the self-generated turbulence driven by the
velocity shear. Nevertheless, a proper analysis of the growth-rates of such
instabilities must be conducted in future work.

Another consequence of the mechanism responsible for turbulence self-generation
is on the statistics of perturbations. Velocity shear is a global process that
occurs in regular magnetized and unmagnetized fluids. The nonlinear evolution of
related instabilities, as the Kelvin-Helmholtz instability, is known to be one
of the main contributors to the energy transfer rate between wavemodes, i.e. the
energy cascade. If the energy cascade in reconnection layers is led by similar
mechanisms, it is straightforward to understand why the statistics observed
resembles those of Kolmogorov-like turbulence, and Goldreich-Sridhar anisotropy
scaling. In other words, our claim is that the turbulent onset and cascade in
reconnection is not different to those found in regular MHD and hydrodynamical
systems.

\section{Conclusions}
\label{sec:conclusions}

We performed numerical modeling of a realistic setup in which magnetic
reconnection can develop turbulence from the initial weak noise of velocity
fluctuations. We analyzed time evolution of several quantities, including
kinetic energy, its distribution among the velocity components, mean vorticity,
its filling factor, the thickness of turbulent region, and we estimated the
reconnection efficiency using two independent methods. Once the turbulence was
developed in a broad region we analyzed its properties using power spectra of
velocity and vorticity, and anisotropy scaling of velocity fluctuations with
respect to the local mean field.

The conclusions coming from these studies can be summarized in several points:
\begin{itemize}
 \item Reconnection is able to develop substantial amount of turbulence from
initially weak noise of velocity fluctuations. We observe growth of the kinetic
energy by over three orders of magnitude.
 \item The generation of rotational motions measured by vorticity is very quick
initially, and reaches saturation after around $5.0 t_A$. The saturation level
of mean vorticity depends on the grid size $h$ and $\beta$, being higher for
models with smaller effective grid size and smaller $\beta$. The filling factor
of vorticity with a threshold $|\vec{\omega}| \ge 10.0$ depends also on $h$ and
$\beta$, however, for model with the largest $h$, the filling factor decays
after reaching a maximum value around $7.0 t_A$, while for the highest
resolution model it constantly grows until the end of simulation. This indicates
that the filling factor of strong rotational motions should be much higher in
realistic systems. In models with $L_z = L_x$ the filling factor of vorticity
grows for all models with different sound speeds, even though the effective grid
size was $h = 1/512$.
 \item The estimated reconnection rates using two independent methods, i.e. the
growth rate of the turbulent region thickness $d\Delta/dt$ and the inflow speed
$V_{in}$ with which the fresh magnetic flux is brought to the system, are
independent or weakly dependent on the resolution for similar $\beta$,
indicating that the reconnection could be fast without necessity of the presence
of external processes, such driven turbulence, or collisionless plasma effects.
The dependence of $V_{in}$ on $\beta$ might be related to additional magnetic
field dissipation in supersonic shocks created within the turbulent region.
 \item The velocity and vorticity present power spectra compatible with
Kolmogorov slope. There is a small bump observed at small scales which could be
explained by the action of reconnection or sort of bottleneck effect.
 \item The anisotropy degree and scaling depends on the $\beta$-plasma
parameter, related to the timescale of decay of the reconnection outflow
interactions to turbulent fluctuations.
 \item The velocity fluctuations generated by reconnection-driven turbulence are
the \cite{GoldreichSridhar:1995} model compatible, however, their statistics are
strongly distorted by reconnection outflows driving the turbulence, especially
for low $\beta$ models, where supersonic motions can be generated. In high
$\beta$ models, the velocity anisotropy follows the $l_\parallel \propto
l_\perp^{2/3}$ scaling. This scaling is also visible at large scales for the
highest resolution model with $\beta \approx 1.0$.
\end{itemize}

\acknowledgments
G.K. acknowledges support from FAPESP (grants no. 2013/04073-2 and 2013/18815-0)
and PNPD/CAPES (grant no. 1475088) through a Postdoctoral Fellowship at
University Cruzeiro do Sul. This work has made use of the computing facilities
of the Laboratory of Astrophysics (EACH/USP, Brazil) and the Academic
Supercomputing Center in Krak\'ow, Poland (Supercomputer Prometheus at ACK
CYFRONET AGH). D.F.G. thanks the Brazilian agencies CNPq (no. 302949/2014-3) and
FAPESP (no. 2013/10559-5) for financial support. A.L. acknowledges the NSF grant
AST 1212096, NASA grant NNX14AJ53G as well as a distinguished visitor PVE/CAPES
appointment at the Physics Graduate Program of the Federal University of Rio
Grande do Norte, the INCT INEspaço and Physics Graduate Program/UFRN. E.T.V.
acknowledges the support the AAS.

\bibliography{references}

\end{document}